\documentclass[12pt]{extarticle}
\usepackage{amsmath}
\usepackage{amsthm}
\usepackage{amssymb}
\usepackage{stackengine}
\usepackage{caption}
\usepackage{comment}
\usepackage{enumitem}
\usepackage{float}
\usepackage{natbib}
\usepackage{subfigure}
\usepackage{xcolor}
\usepackage{titling}
\usepackage{graphicx}
\usepackage[letterpaper]{geometry} 
\usepackage{tikz}
\usetikzlibrary{patterns.meta}
\usetikzlibrary{backgrounds}
\usepackage{array}
\usepackage{authblk}
\usetikzlibrary{positioning, fit, calc}

\usepackage[hidelinks]{hyperref}

\ifdefined\REVIEW
\newcommand{\new}[1]{\textcolor{blue}{#1}}
\else
\newcommand{\new}[1]{#1}
\fi

\theoremstyle{definition}

\title{Modeling Discrimination with Causal Abstraction}

\author{
\stackunder{Milan Mossé$^*$}{{\small {\sl UC Berkeley}}}\,\, \stackunder{Kara Schechtman$^*$}{{\small {\sl Princeton}}} \,\,\stackunder{Frederick Eberhardt}{{\small {\sl Caltech}}}\,\, \stackunder{Thomas Icard}{{\small {\sl Stanford}}}
}
\date{
\small August 2025
}

\begin{document}

\maketitle
\def\thefootnote{*}\footnotetext{Joint first author.}\def\thefootnote{\arabic{footnote}}

\begin{abstract}

    A person is directly racially discriminated against only if her race caused her worse treatment. This implies that race is an attribute sufficiently separable from other attributes to isolate its causal role. But race is embedded in a nexus of social factors that resist isolated treatment. If race is socially constructed, in what sense can it cause worse treatment? Some propose that the perception of race, rather than race itself, causes worse treatment. Others suggest that since causal models require \textit{modularity}, i.e. the ability to isolate causal effects, attempts to causally model discrimination are misguided.

    This paper addresses the problem differently. We introduce a framework for reasoning about discrimination, in which race is a high-level \textit{abstraction} of lower-level features. In this framework, race can be modeled as itself causing worse treatment. Modularity is ensured by allowing assumptions about social construction to be precisely and explicitly stated, via an alignment between race and its constituents. Such assumptions can then be subjected to normative and empirical challenges, which lead to different views of when discrimination occurs. By distinguishing constitutive and causal relations, the abstraction framework pinpoints disagreements in the current literature on modeling discrimination, while preserving a precise causal account of discrimination.

\end{abstract}

\newpage

\section{Introduction}

Direct racial discrimination against a person occurs only if she is treated worse than others in some respect, and this is explained by her race \citep{KlemThomsen2018}. When explanation is understood causally, we arrive at the notion of

\begin{quote}
    \textit{Causal Discrimination:} A person is directly racially discriminated against only if she is treated worse than others, and this is caused by her race.\footnote{\new{We focus here just on direct discrimination, but some have argued that race causing worse treatment is a necessary condition for all objectionable forms of discrimination, including both direct and indirect discrimination \citep{loi2023would}.}}
\end{quote}

This notion of discrimination undergirds the design of \textit{audit studies}, in which experimenters detect institutional discrimination using techniques that resemble randomized controlled trials. In a now-famous correspondence study, resumes were randomly assigned Black- or white-sounding names and sent out to prospective employers. White names received 50 percent more callbacks for interviews, and callbacks were more responsive to resume quality for white names than for Black ones \citep{bertrand2004are}. In other words, the study found that intervening on race-coded names changed interview callbacks, irrespective of resume quality; in this sense, race caused worse treatment. The causal notion of discrimination then justifies the researchers' conclusion that their study revealed racial discrimination in the labor market. This kind of audit study is widely considered a gold standard for detecting discrimination, with social scientists undertaking increasingly larger-scaled, more ambitious, and more methodologically complex experimental designs \citep{gaddis2018introduction}. 

The causal notion of discrimination popularized by audit studies applies equally to other protected attributes, like gender or religion, and is invoked across a variety of settings. For example, causal discrimination is invoked by algorithmic fairness criteria \citep{kilbertus2017avoiding,kusner2017counterfactual,barocas-hardt-narayanan,Plecko}\footnote{These causal criteria hold the promise of escaping some of the impossibility theorems \citep{kleinberg2017inherent, chouldechova2017fair,barocas-hardt-narayanan} that plague merely probabilistic or ``correlational'' notions of discrimination \citep{glymour2019measuring,beigang2023reconciling}. More generally, it is suggested that causal criteria appropriately cash out the notion of explanation relevant to algorithmic fairness \citep{hedden2021statistical}. Of course, causal analysis requires substantive causal assumptions, but reasoning with such assumptions is no harder, from a computational complexity perspective, than probabilistic reasoning \citep{mosse2022causal, vanDerZander2023}, and theorists have developed methods for protecting against causal discrimination in the face of uncertainty about the underlying causal structure \citep{russell2017worlds}.}, and in the U.S. law \citep{kohlerhausmann2018eddie}, as in a recent Supreme Court decision:

\begin{quote}
    Title VII \dots prohibits employers from taking certain actions ``because of'' sex. \dots Title VII's ``because of'' test incorporates the ``simple'' and ``traditional'' standard of but-for causation. \dots That form of causation is established whenever a particular outcome would not have happened ``but for'' the purported cause. \dots In other words, a but-for test directs us to change one thing at a time and see if the outcome changes. If it does, we have found a but-for cause. \citep{bostock_clayton_county}
\end{quote}

\noindent Here, the Court explicitly invokes the causal notion of discrimination: discrimination occurs when, were sex different, the outcome would have been different as well.

In sum, the commonsense idea that discrimination occurs only when a protected attribute causes worse treatment is invoked across the social sciences, computer science, and U.S. law. But even assuming we have settled on a theory of when and why attributes such as race and gender should be protected from discrimination (e.g. \citealt{hellman2008discrimination, Kolodny2023}), the notion of causal discrimination is difficult to spell out in a precise and plausible way. Indeed, on the counterfactual understanding of causation \citep{lewis1973causation,woodward2003a,woodward2003b,woodward2005making}, to determine whether race caused worse treatment, we must consider whether an individual's treatment would have been worse, had their race and \textit{only} their race been different;\footnote{We focus on the framework of structural causal models rooted in the work of \cite{Pearl2000-PEAC-2}, given that it is used in definitions of algorithmic fairness \citep{kilbertus2017avoiding,kusner2017counterfactual}, and that its ability to accommodate racial discrimination has already been subject to criticism \citep{kohlerhausmann2018eddie,hu2020sex,hu2024} and defense \citep{weinberger22}. For discussion of analogous difficulties for the potential outcomes framework, see \citet{glymour2014race, marcellesi2013race, greiner2011causal, sen2016race}. For an equivalence between structural causal models and potential outcomes frameworks, see \cite{ibeling2024comparing}.} this is precisely the test suggested by audit studies, algorithmic fairness criteria, and the U.S. Supreme Court.

However, it is widely held that race, as well as other attributes protected from discrimination like gender, religion, and disability,  are \textit{socially constructed}: they help to sustain and are  influenced by social norms concerning individuals' appearance, wealth, location, education, criminal history, and more \citep{Haslanger2000, haslanger2012resisting,taylorRacePhilosophicalIntroduction2022,mills1998blackness}. This raises the worry that there is no robust metaphysical boundary between what \textit{constitutes race} (and thus should be varied, when testing racial discrimination) and what is \textit{causally distinct from race} (and thus should be controlled for, when testing racial discrimination).% For instance, in the resume audit study, why shouldn't an auditor vary features of individuals like educational history in addition to their names, for example by changing university education from ``Harvard'' to ``Howard,'' if such features are bound up in the social construction of race? MM: In comments on the draft, Christian says that this sentence is a little distracting, and can make the reader think that our focus is on what we later call intervention identification, rather than the modularity problem

An influential line of recent work argues that standard causal frameworks for modeling discrimination make it impossible to give a satisfactory answer to this worry, because they force us to postulate an implausible metaphysical boundary between race and the context in which it is socially embedded \citep{kohlerhausmann2018eddie, hu2020sex, hu2024}. Race can cause worse treatment in a causal model only if its effects are \textit{modular}: one can perform an intervention \textit{only} on race and trace the resulting effects to race. But if race is socially constructed, it seems to violate this requirement of modularity: in intervening on race, one inevitably intervenes on its constituents as well. For example, because race is partly constituted by norms surrounding people's names, education, and employment history, interventions on these attributes and on race are intertwined. %bound up with each other. 
If we change someone's race from Black to white while attempting to hold fixed ``Jamal'' as their name or ``Howard University'' as part of their educational history, the social resonance of these attributes changes; interventions on race, on names, and on education are intertwined. The modularity requirement then seems to forbid a causal model from including variables corresponding to race, names, education, and so on, when precisely such a model would be needed to explain how manipulations on names reveal effects of race in the resume audit study.

More generally, causal models of discrimination face a \textit{modularity problem:}
\begin{enumerate}
    \item 
    \textit{Causal discrimination:} A person is directly racially discriminated against only if she is treated worse than others, and this is caused by her race.
    \item 
    \textit{Modularity assumption:} Race is a cause in a causal model only if its effects are modular.
    \item 
    \textit{Social construction:} Race is socially constructed, so its effects violate modularity.
    \item 
    Therefore, causal models cannot be used to model racial discrimination.
\end{enumerate}

This worry has prompted many to defend causal models, typically by arguing that causal models of discrimination do not require modularity for race, but rather for some distinct feature associated with race. For example, \cite{greiner2011causal} propose that racial discrimination occurs if others' \textit{perception} of race causes worse treatment (cf. \citealt{Singh2023}), \cite{sen2016race} propose that race has several constituents and that racial discrimination occurs if any \textit{constituent} causes worse treatment, and \cite{weinberger22} proposes that racial discrimination occurs if \textit{signals} of race cause worse treatment. We argue that the modularity problem is not adequately addressed by these proposals.

Meanwhile, those who developed the modularity problem conclude from it that we must think of race as explanatory in some sense that cannot be represented by causal models \citep{kohlerhausmann2018eddie, hu2020sex, dembroff2020taylor, hu2024}. Like the alternatives of skepticism \citep{holland1986statistics} or agnosticism \citep{tolbertcausal} that race is a cause, this amounts to a rejection of the notion of causal discrimination apparently deployed by audit studies, algorithmic fairness criteria, U.S. law, and by those describing their lived experience of discrimination \citep[pp.45-46, 53-55]{cherryMakingItCount2018}. At points these authors suggest that discrimination claims are backed by constitutive explanation, rather than causal explanation (\citealt{hu2020sex}, p.10; \citealt{dembroff2020taylor}, p.9). But we will argue that this proposal is best understood as pointing to a distinction between two kinds of causal discrimination, and that this distinction, as well as the relations between race and its constituents, are well-represented by causal models.\footnote{A third line of response would be to claim that protected attributes like race and gender are not socially constructed, and instead adopt \textit{skepticism} that race exists \citep{appiah1995, zack1993} or \textit{racial population naturalism}, which holds that ``it is possible that genetically significant biological groupings could exist that would merit the term races . . . [though] there is no set of genetic or other biological traits that all and only all members of a racial group share that would then provide a natural biological boundary between racial groups.'' \citep{sep-race} To address the modularity problem, one would need to adopt such views regarding \textit{every} attribute which is a potential basis for wrongful discrimination, e.g. gender, gender identity \citep{Stoljar1995,Jenkins2016}, racial identity \citep{Appiah1998}, and race as it is commonly perceived and conceptualized. The abstraction model can be used to model (social or natural) constituents of any of these attributes. Since the modularity problem openly assumes that protected attributes are socially constructed, we prefer a reply to it which is consistent with social construction.}

Crucially, both of these responses to the modularity problem share what we can call the

\begin{quote}
    \textit{Single-Level Assumption:} When causally modeling some phenomenon, one must place all of the relevant variables in a single causal model, which describes the situation at a single level of abstraction.
\end{quote}

\noindent In this paper, we propose a simple framework for modeling discrimination which jettisons the Single-Level Assumption, by modeling race as a high-level abstraction of lower-level attributes. We introduce the

\begin{quote}
    \textit{Abstraction Model for Discrimination:} A person's race is constituted by low-level natural and social attributes and relations. Race causes an outcome if a change in race, instantiated by a change in the low-level attributes or relations constitutively aligned with race, causes a change in outcome. This is the sense in which race causes worse treatment, when racial discrimination occurs.
\end{quote} 

While we develop the framework by applying it to racial discrimination in audit studies, it applies equally to other protected attributes, like gender, and in other settings, including algorithmic fairness and the U.S. law. We show that this model dissolves the modularity problem: in our model, the relationship between race and its constituents remains non-modular, in the sense that one cannot intervene on race without intervening on its constituents. But as we explain, this is a benign (indeed intended) feature of abstraction, not an obstacle to causal modeling: when one attribute is simply a higher-level abstraction of other, lower-level attributes, one should expect changes at one level to correspond to changes at the other.

The abstraction framework addresses the modularity problem, but the use of abstraction does not obviate the need to make normative judgments about the relevant counterfactual contrasts in testing the effects of social kinds. Instead, the abstraction framework provides a way of making these judgments precise and explicit, primarily via an alignment between higher-level protected attributes and their lower-level constituents. In this way, the abstraction framework offers the flexibility to encode different assumptions about how discrimination should be measured, which lead to different verdicts about whether discrimination has occurred. Made precise and explicit, these assumptions can then be subjected to further normative and empirical challenges. 

Indeed, after introducing the abstraction framework  (Section~\ref{sec:abstraction}) and addressing the modularity problem (Section~\ref{sec:modularity}), we use the framework to distinguish further challenges for causal models of discrimination (Section~\ref{sec:further challenges}). In particular, we address the worry that interventions which purport to test the discriminatory effects of race in fact reveal quantities which are in some sense irrelevant to discrimination \citep{hu2020sex}. We then turn to the objection that causal models cannot capture discrimination claims, because discrimination involves constitutive, rather than causal, explanation \citep{dembroff2020taylor}. \new{We argue that this is not an objection to causal models of discrimination as such, because it can be precisely stated within the framework of causal abstraction. In this framework, the objection motivates a distinction between two notions of causal discrimination: the effects of a social kind can be causally explained by a \textit{particular individual's membership in the kind}, or by the \textit{norms that constitute} the social kind (Section~\ref{sec:two notions}). In this way, the abstraction framework allows us to pinpoint disagreements in the current literature on modeling discrimination, while preserving a precise causal account of discrimination.}

\section{Abstraction and discrimination}\label{sec:abstraction}

The social construction of protected attributes like race implies that an individual's race is not a feature on par with other natural or social attributes and relations, but rather defined in terms of such ``lower-level'' attributes and relations. While such multi-level representation is not part of a standard causal model, causal abstractions can represent causal systems at multiple levels of granularity. This section will introduce the framework of causal abstraction and then show how it can be used to model discrimination, using the example of audit studies to illustrate.

\subsection{Introducing abstraction}\label{sec:introducing abstraction}

Consider an experiment in which a bird is trained to peck at objects of any shade of red \citep{Yablo1992,woodwardExplanatoryAutonomyRole2021}. When the bird pecks at some red object, say a crimson one, we can make two causal claims about it, at different levels of abstraction: that it pecks at the object because it is \textit{crimson}, or that it pecks at an object because it is \textit{red}. Each of these causal claims suggests an associated causal model, one in which perception of a fine-grained color (e.g., crimson, scarlet, cyan, turquoise) causes pecking behavior, and another in which perception of a coarse-grained color (e.g., red, blue) causes pecking behavior. Intuitively, the coarse-grained model is an abstraction of the lower-level, fine-grained one, because (a) the coarse-grained model is \textit{simpler}, and (b) the models are \textit{causally equivalent}: changing the color from red to blue in the coarse-grained model has the same effect as changing the color from crimson to cyan in the low-level model. As we now explain, in the framework of causal abstraction, the notion of ``simplifying a model'' is provided by a relationship between them, called an \textit{alignment}, while the notion of causal equivalence is provided by a feature of alignments, called \textit{causal consistency.}

\begin{figure}[ht]
    \centering
    \begin{tikzpicture}[->, >=stealth,shorten >=1pt,auto,node distance=2cm,thick]
        % Original nodes and arrow
        \node[circle, draw, minimum size=6em] (FineColor) {\textsf{Fine}};
        \node[circle, draw, minimum size=6em, right=of FineColor] (Pecking) {\textsf{Pecking}};
        \path (FineColor) edge[->] node {} (Pecking);

        \node at (0,-.4) {\scriptsize{}$\{\textnormal{crimson, cyan,}$};
        \node at (0,-.7) {\scriptsize{}$\textnormal{ turquoise...}\}$};

        \node at (0,3) {\scriptsize{}$\{\textnormal{red, blue}\}$};

        \node at (4.5,-.5) {\scriptsize{}$\{\textnormal{yes, no}\}$};
        
        % Copied nodes and arrow, moved down
        \node[circle, draw, minimum size=6em, above=10mm of FineColor] (CoarseColor) {\textsf{Coarse}};
        \node[circle, draw, minimum size=6em, right=of CoarseColor] (Pecking2) {\textsf{Pecking}};
        \path (CoarseColor) edge[->] node {} (Pecking2);

        \node at (4.5,3) {\scriptsize{}$\{\textnormal{yes, no}\}$};
        
        \path (FineColor) edge[dotted] node[midway,left] {{$\tau_{\text{color}}$}} (CoarseColor);
        \path (Pecking) edge[dotted] node[midway,left] {{$\tau_{\text{color}}$}} (Pecking2);

        % Labels for models
        \node[left=10mm of FineColor] (lowModel) {{Low-level model}};
        \node[left=10mm of CoarseColor] (highModel) {{High-level model}};
    \end{tikzpicture}

    \captionsetup{font=small}
    \caption{Color as an abstraction. Dotted arrows represent an alignment $\tau_{\text{color}}$ between causal models, which maps values of the \textsf{Fine} color variable (crimson, scarlet) to values of the \textsf{Coarse} color variable (red). Solid arrows express causal relationships, e.g.\ that intervening on the value of the \textsf{Coarse} color variable can change the value of the \textsf{Pecking} variable.
    }
    \label{figure:color as abstraction}
\end{figure}

The fine-grained and coarse-grained models are depicted in Figure~\ref{figure:color as abstraction}. The high-level model contains two variables: $\textsf{Coarse}$, which represents the bird's perception of a coarse-grained color, and $\textsf{Pecking}$, which represents whether the bird pecks. The causal arrow implies that $\textsf{Pecking}$ is an effect of $\textsf{Coarse}$, and thus is some function %$f_{\textsf{coarse}}$ 
of $\textsf{Coarse}$, perhaps together with random noise (included to capture variation among birds' pecking behavior not explained by object color). The low-level model is almost exactly the same, except it models pecking behavior as a function of fine-grained color $\textsf{Fine}$ rather than coarse color. \new{We'll now introduce the relevant notion of a causal model slightly more formally, and then explain alignments and causal consistency.}

In general, a \textit{structural causal model} $\mathcal{M}$ contains variables $V_1,V_2,V_3\dots$, where the value of each variable $V$ is defined by a function $f_{V}$ of the values of other variables, and perhaps some random noise, to capture variation not explained by model variables 
(see, e.g., \citealt{Pearl1995,peters2017elements,bareinboimHierarchy}.) The effect of a manipulation on a variables is modeled as an \textit{intervention} $\text{do}(V = v)$, which replaces the function $f_V$ defining $V$ with a fixed value $v$. Because other variables may be defined as functions of $V$, the intervention $\text{do}(V = v)$ percolates through the model, changing the likelihood that other variables take various values; this is the effect of the intervention.

Consider two models $\mathcal{M}_{\text{low}}$ and $\mathcal{M}_{\text{high}}$. An \textit{{alignment}} $\tau$ is a function which maps values of variables in the low-level model $\mathcal{M}_{\text{low}}$ to values of variables in the high-level model $\mathcal{M}_{\text{high}}$. Intuitively, an alignment may be thought of as a \textit{simplification} of the low-level model: it merges different values of low-level variables into individual values of high-level variables, and may ignore some low-level variables entirely \citep{Chalupka,Rubinstein2017,beckersApproximateCausalAbstraction2019a,geiger2024causalabstractiontheoreticalfoundation}. \new{In the above example, the alignment $\tau_{\text{color}}$ maps crimson and scarlet to red, while mapping cyan and turquoise to blue. More generally, the alignment $\tau$ need not be injective, and as a result, interventions on high-level variables will often be ambiguous between several different interventions on low-level ones. (We discuss this issue further in Section~\ref{sec:further challenges}.)}

So-defined, alignments are highly permissive: strictly speaking, any model can be aligned with any other. As we now explain, to define an abstraction, an alignment must also ensure that interventions in the low-level model and interventions in the high-level model are ``causally equivalent.'' Formally, an alignment establishes causal equivalence between two models when it is \textit{causally consistent.} To check whether $\tau_{\text{color}}$ is causally consistent, we first perform an intervention on color at the low level, and then transform the effect it has on pecking to the high level via the alignment. We then check that this has the same result as first transforming the intervention via the alignment, and then intervening at the high level. This condition is in fact met by the alignment $\tau_{\text{color}}$, because intervening on fine-grained color by changing it from crimson to cyan and then checking whether the bird pecks has the same result as intervening on coarse-grained color by changing it from red to blue and then checking whether the bird pecks. By mapping more general colors to corresponding more specific shades, $\tau_{\text{color}}$ tracks a real metaphysical relationship between a color and its many shades, to which pecking behavior is insensitive. More generally, an alignment $\tau$ between two models is causally consistent if intervening on variables in the low-level model $\mathcal{M}_{\text{low}}$ and then aligning into the high-level model $\mathcal{M}_{\text{high}}$ is the same as first aligning into the high-level model and intervening there. Given such an alignment, we say that $\mathcal{M}_{\text{high}}$ is an {\textit{abstraction}} of $\mathcal{M}_{\text{low}}$. (See e.g. Def. 3 of \citealt{Rubinstein2017} or Def. 16 of \citealt{geiger2024causalabstractiontheoreticalfoundation}.)

Causal consistency is a stringent requirement, requiring perfect correspondence between low-level and high-level interventions. We introduce it for simplicity, but it can be relaxed in various ways (see, e.g., \citealt{beckersApproximateCausalAbstraction2019a,rischelCompositionalAbstractionError2021,geiger2024causalabstractiontheoreticalfoundation}). Most simply, where the values of several lower-level variables $V_1,....,V_n$ are aligned via $\tau$ with the values of a single high-level variable $V$, we can say that $\tau$ is \textit{approximately} causally consistent if the \textit{average} of the effects of $V_1,...,V_n$ on some outcome is very close to almost all of the effects of $V_1,...,V_n$ on that outcome, taken individually (cf. \citealt[Appendix D1]{Plecko}).%\footnote{E.g., suppose that are 10 shades of red $r_1,...,r_{10}$, and that all but $r_{10}$ prompts pecking behavior. Then an alignment $\tau$ between these shades and the color red is not causally consistent, because some intervention which sets the color to red cause pecking behavior, while others do not. Nonetheless, $\tau$ is approximately causally consistent, because (1) 90\% of all interventions which set color to red result in pecking behavior, so the \textit{average effect} is pecking behavior; and (2) almost all of the lower-level shades of red have this same effect. In other words, $\tau$ is approximately causally consistent because the average of the effects of the low-level attributes aligned with red is very close to individual effects of almost all of these attributes. (The notions of ``almost all'' and ``very close'' can be made precise with constants and metrics that measure the distances among variables' values and among their distributions on those values. One may in addition use a weighted average, rather than a simple mean, in computing the average effect of low-level interventions.)}

\subsection{Modeling social constructions with abstraction}\label{sec:abstraction:social_construction}

Because any metaphysical relationship between two attributes that necessarily co-vary can be captured by an alignment, causal abstraction is a flexible framework. For example, it can be used to express logical relations, determinate-determinable relations, and constitutive relations. In this section, we explain how to use abstraction to capture the constitutive relations posited by a social constructionist theory of race. The next section provides a formal interpretation of the causal discrimination claims tested by audit studies.

\new{We begin with a gloss on social constructionism. Our understanding of practically any kind is of course shaped partly by contingent features of our linguistic practices---consider our criteria distinguishing hurricanes from typhoons. But here, we focus on kinds that are \textit{socially constructed}, in either of the following senses \citep[§1.3]{sep-social-construction-naturalistic}:}

\begin{enumerate}[label=(\roman*)]
    \item 
    \new{A kind is \textit{causally} socially constructed when the causal role and characteristic features of members of the kind depend on how the kind is classified \citep[p. 98]{haslanger1995ontology}.}
    \item 
    \new{A kind is \textit{constitutively} socially constructed when individual membership in the kind is largely constituted, as a matter of convention, by social facts, for example facts about social relations or social status \citep[p. 20]{haslanger2017gender}.}
\end{enumerate}

\noindent \new{Consider  Haslanger's~\citeyearpar[pp. 6-7]{haslanger1995ontology} view of the kind \textit{woman}:}
\begin{quote}
    \new{[An individual] $S$ is a woman iff $S$ is systematically subordinated along some dimension (economic, political, legal, social, etc.), and $S$ is ``marked'' as a target for this treatment by observed or imagined bodily features presumed to be evidence of a female’s biological role in reproduction.}
\end{quote}

\noindent \new{On this view, \textit{woman} is socially constructed in both senses. It is \textit{constitutively} socially constructed, because relations of subordination are part of what constitute an individual’s gender, and it is \textit{causally} socially constructed, because this characteristic feature of the kind is explained by the way individuals are classified (or ``marked''). The view that individuals’ sex and gender attributes are constitutively socially constructed is attractive, because it is thought to make good sense of transgender and genderqueer identities (Jenkins~\citeyear{Jenkins2016}; Dembroff et al.~\citeyear[p.4]{dembroff2020taylor}), as well as socially-determined variation in how intersex individuals are sexed and gendered \citep[pp. 726-7]{asta2013social}.}

\new{On a social constructionist theory of race, to belong to a certain race is to have a collection of attributes that are imbued with racial meaning as part of a process of ``racial formation'' \citep[pp. 4-7]{omiTheoryRacialFormation2014}, which causally socially constructs race. For ``thin'' social constructionists, race is causally but not constitutively socially constructed: racial attributes are based only on individuals’ appearances and ancestry \citep[pp. 11-13]{mallonRaceNormativeNot2006}. But ``thick'' social constructionism---which motivates the modularity problem (Hu~\citeyear[§4]{hu2024}; Hu and Kohler-Hausmann~\citeyear[pp. 249-51]{huWhatPerceivedWhen2024})---holds that individuals’ races are also constituted by social facts. For example:}
\begin{itemize}
    \item 
    \new{Taylor~\citeyearpar[pp. 139-144]{taylorRacePhilosophicalIntroduction2022} argues that the constituents of an individual’s race are their appearances, ancestry, and social position in the set of social conditions statistically correlated with appearances and ancestry, including employment, insurance, and housing conditions.}
    \item 
    \new{Haney-López~\citeyearpar[p. 39]{haney-lopezianSocialConstructionRace1994} argues that an individual’s race is defined by a combination of ``chance, context, and choice,'' with chance referring to ``morphology and ancestry,'' context referring to ``the contemporary social setting,'' and choice ``the quotidian decisions of life.''}
    \item 
    \new{Mills~\citeyearpar[pp. 51-55]{mills1998blackness} argues that racial self-and-other identifications draw on seven criteria: bodily appearance, ancestry, self-awareness of ancestry, public awareness of ancestry, culture, experience, and self-identification.}
\end{itemize}

\new{Even those who endorse constitutive social constructionism about gender might object to a thick constructionist view of race. Indeed, one might worry that this view puts too many attributes in the constitutive basis of race, suggesting individuals could change their race just by  changing their ``superficial'' attributes---for example, attending a historically Black university or moving to a predominantly-white neighborhood. While a detailed discussion of this worry would take us beyond the scope of this paper, here are three points in response. First, views that limit race’s constitutive basis to appearances and ancestry struggle to describe the limited cases where an individual’s race is in  their control---such as in cases of passing as another race \citep[pp. 41-66]{mills1998blackness}, or racial categorization changing with location \citep[p. 7]{griffithSocialConstructionGrounding2018}. Second, including features in the constitutive basis of race that individuals can control need not imply that changing these features always suffices to change race. To the extent that people have control over their race, it is heavily contextually constrained. Given the vast array of social roles and other attributes that go into race, thick social constructionism does not imply that individuals can simply ``switch'' their races on a whim (Haney-Lopéz~\citeyear[pp. 47-8]{haney-lopezianSocialConstructionRace1994}, Taylor~\citeyear[p. 112]{taylorRacePhilosophicalIntroduction2022}). Third, even choices of affiliation that do not change race, such as a white person attending a historically-Black university or a Black person living in a predominantly white neighborhood, are acts imbued with racial meaning, and so arguably belong to the constitutive basis of race \citep[pp. 49-50]{haney-lopezianSocialConstructionRace1994}.}

\new{Abstraction provides a way to formalize the claim that attributes like race and gender are \textit{constitutively} socially constructed. The relations between race and its constituents induce an alignment $\tau_{SC}$ that specifies how an individual's appearance, ancestry, and social conditions map onto their race (Figure~\ref{figure:race+resume as abstractions}). We focus on constitutive (rather than causal) social constructionism for two reasons. First, as suggested above, this is the form of social constructionism that motivates the modularity problem raised by Hu and Kohler-Hausmann. Second, audit studies often claim to establish conclusions about the effects of an \textit{individuals'} attributes (as opposed to, say, the effects of \textit{perceptions} of these attributes or of \textit{population-level attributes}).}\footnote{\new{We underscore that even if race is in some sense \textit{grounded} or \textit{realized} by various further attributes (cf. \cite{griffithSocialConstructionGrounding2018}) rather than \textit{constituted} by them, abstraction can still be used to represent the metaphysical relationship between race and these further attributes: what matters is that various attributes of an individual are mapped to their race. See \cite{dasguptaConstitutiveExplanation2017} for an argument that the grounding relation is best understood as constitution.}}

\begin{figure}[H]
    \centering
    \begin{tikzpicture}[->, >=stealth, shorten >=1pt, auto, node distance=2cm, thick]
        % Race
        \node[circle, draw, minimum size=5em,pattern={Lines[angle=45,distance=8pt,line width=1pt,]}, pattern color=blue!70, fill opacity=100] (Race) {\textsf{Race}};

        % Resume
        \node[rectangle, dashed, draw, minimum size=7em, right = 90mm of Race] (Resume) { };
        \node[below =1mm of Resume] {Resume};

        \node[draw, circle, below=-15mm of Resume,
     pattern={Dots[distance=4pt,radius=1pt]}, pattern color=orange!70] (R1) {\textsf{Work}};

        \node[draw,circle,pattern={Dots[distance=4pt,radius=1pt]}, pattern color=orange!70] at ([shift={(-.7cm,1.25cm)}]R1) {\textsf{Name}};
        
        \node[draw,circle,pattern={Dots[distance=4pt,radius=1pt]}, pattern color=orange!70] at ([shift={(.7cm,1.25cm)}]R1) {\textsf{Edu}};
        
        \coordinate (Mid) at ($(Race)!0.5!(Resume)$);

         % All attributes
        \node[rectangle, dashed, draw, minimum size=12em, below=-2mm of Mid] (features) { };
        
        \node[below =1mm of features] {All attributes};
        
        \node[draw, circle, below=-14.4mm of features,pattern={Lines[angle=45,distance=8pt,line width=1pt,]}, pattern color=blue!70, fill opacity=10] (F1) {\textsf{Work}};
    
    \node[draw, circle, below=-14.4mm of features,pattern={Dots[distance=4pt,radius=1pt]}, pattern color=orange!70] (F1) {\textsf{Work}};

        \node[draw,circle,pattern={Lines[angle=45,distance=8pt,line width=1pt,]}, pattern color=blue!70, fill opacity=10] at ([shift={(-1.25cm,1cm)}]F1) {\textsf{Name}};

    \node[draw,circle,pattern={Dots[distance=4pt,radius=1pt]}, pattern color=orange!70] at ([shift={(-1.25cm,1cm)}]F1) {\textsf{Name}};

        \node[draw,circle,pattern={Lines[angle=45,distance=8pt,line width=1pt,]}, pattern color=blue!70, fill opacity=10] at ([shift={(-1.6cm,3.5cm)}]F1) {\textsf{Edu}};
    
    \node[draw,circle,
    pattern={Dots[distance=4pt,radius=1pt]}, pattern color=orange!70] at ([shift={(-1.6cm,3.5cm)}]F1) {\textsf{Edu}};

    \node[draw,circle] at ([shift={(1.6cm,.25cm)}]F1) {\textsf{\dots}};
        
        \node[draw,circle,pattern={Lines[angle=45,distance=8pt,line width=1pt,]}, pattern color=blue!70, fill opacity=10] at ([shift={(.5cm,1.5cm)}]F1) {\textsf{Family}};

        \node[draw,circle] at ([shift={(-.7cm,2.4cm)}]F1) {\textsf{Age}};
        
        \node[draw,circle,pattern={Lines[angle=45,distance=8pt,line width=1pt,]}, pattern color=blue!70, fill opacity=10] at ([shift={(1.2cm,3.2cm)}]F1) {\textsf{Appear}};

        \node at (2.2,-.6) {$\tau_{\text{SC}}$};
        \path (features) edge[dotted] node[midway,right] {} (Race);

        \node at (9,-.6) {$\tau_{\text{audit}}$};
        \path (features) edge[dotted] node[midway,left] {} (Resume);

    \end{tikzpicture}

    \captionsetup{font=small}
    \caption{Race and resume as abstractions. The alignment $\tau_{\text{SC}}$ comes from social constructionist theories of race, and the alignment $\tau_{\text{audit}}$ is used by audit studies. The dotted box labeled ``all attributes'' contains all attributes of an individual, some of which are abstracted into race \new{and marked with blue lines going northeast} (e.g., appearance, name, work, family history, education in the diagram), and others of which are discarded by the alignment $\tau_{\text{SC}}$ (e.g. age). The dotted ``Resume'' box represents a collection of variables \new{from all attributes}, including work history, education, name, and so on; \new{these are marked with orange dots}. \new{The ellipses above indicate that many attributes which might belong on a resume or in the constitutive basis of race are not pictured.}}
    \label{figure:race+resume as abstractions}
\end{figure}

This already involves a substantive modeling assumption. One might hold that it is impossible, even in principle, to specify how exactly race is constructed from lower-level social and natural attributes of individuals. For example, one might claim that race, socioeconomic class, and gender are all mutually constituted by each other, rather than by a more fine-grained basis of lower-level attributes.\footnote{For arguments that would make room for this possibility, e.g., by rejecting the traditional view that constitutive relations are asymmetric and irreflexive, see \cite{barnes2018symmetric} and \cite{Jenkins2016}, respectively. See also \cite{bright2016causally} and \cite{wang2022towards} for further discussion about formally modeling intersectionality.} However, we will set this possible worry aside, because (1) this claim is potentially consistent with the abstraction model, insofar as we may model race as abstracting socioeconomic status and gender in one context, and model gender as abstracting socioeconomic status and race in another; and (2) we assume that social constructionist concerns regarding the possibility of causal models of discrimination do not rest on thorough-going skepticism that a social construction of race in terms of lower-level attributes could ever be articulated in principle. \new{Others might disagree with thick social constructionism entirely; as we later address (Section \ref{sec:modularity:skepticism}), the model is also useful for those who hold thin social constructionist views and those with non-social constructionist views.}

Suppose then that social constructionism induces an alignment $\tau_{\text{SC}}$ between race and the attributes from which it is socially constructed. Because, as the social constructionist suggests, this alignment tracks a genuine metaphysical relationship, we should expect the alignment to be (at least largely) causally consistent: if the effects of race diverged greatly from those of its constituents, this would suggest that $\tau_{\text{SC}}$ had been misspecified. So we can assume that the alignment $\tau_{\text{SC}}$ specifies a causally consistent abstraction.

The simple assumption that race abstracts lower-level attributes in this way already licenses answers to counterfactual questions about race. For instance, it makes sense to ask what would happen were one to intervene on a person's race in a specific way. This is just to ask what would happen, were one to intervene on the lower-level attributes of an individual in such a way that their race would now be different, according to the social construction of race, as defined by the alignment $\tau_{\text{SC}}$.

\subsection{Modeling audit studies with abstraction}

We can use the assumption that race is socially constructed via $\tau_{\text{SC}}$ to propose a causal model of audit studies. Audit studies select a strict subset of an individual's features, namely those appearing on their resume, purportedly ``screening off'' the effects of all other attributes of the individual on the outcome. In the terminology of abstraction, this defines a second alignment $\tau_{\text{audit}}$, which discards any attributes not appearing on the resume and trivially aligns all attributes appearing on the resume with themselves (Figure~\ref{figure:race+resume as abstractions}). The resume is assumed to include all attributes (that is, all variables in ``All attributes'') that exert a causal effect on interview status. In this way, $\tau_{\text{audit}}$ is by construction a ``lossless'' (perfectly consistent) abstraction of ``All attributes'' when it comes to interview status.

% Audit studies test for a causal relationship between resume attributes and interview status. The resume is assumed to include all attributes (that is, all variables in ``All attributes'') that exert a causal effect on interview status. In this way, $\tau_{\text{audit}}$ is by construction a ``lossless'' abstraction of ``All attributes'' when it comes to interview status. The aim in audit studies is to draw conclusions about the effects of race on interview status.
% MM: Combining with the previous paragraph

 We can use the alignments $\tau_{\text{SC}}$ and $\tau_{\text{audit}}$ to state precisely the quantity of interest in common audit studies. A typical audit study fixes two resumes, $\text{Resume}_1$ and $\text{Resume}_2$, which differ only in their names. Because resumes only contain a few pieces of information about an individual, many different people with different attributes could possess the same resume. Where a ``person'' refers to a maximally descriptive set of attributes, let $\mathcal{A}_1$ be the set of people with resume $\text{Resume}_1$, and let $\mathcal{A}_2$ be the set of people with resume $\text{Resume}_2$. (In other words, $\mathcal{A}_1$ is the set of people that $\tau_{\text{audit}}$ aligns with $\text{Resume}_1$, and $\mathcal{A}_2$ is the set of people that $\tau_{\textsf{audit}}$ aligns with $\text{Resume}_2$.)

 % NOTE FOR REVISION: We could relax causal consistency in this section to approximate causal consistency, and replace equality with \approx

The racial compositions of $\mathcal{A}_1$ and in $\mathcal{A}_2$ will tend to differ, especially given that the names appearing on the resumes $\text{Resume}_1$ and $\text{Resume}_2$, like Greg and Jamal, are highly correlated with race. Let $\text{Race}_1$ indicate the race most common in $\mathcal{A}_1$ and let $\text{Race}_2$ indicate the race most common in $\mathcal{A}_2$.\footnote{We align resumes to the most common race in the population for simplicity, but other ways of aligning resumes to races are certainly possible, such as probabilistically aligning $\text{Resume}_1$ to $\text{Race}_1$ and $\text{Race}_2$ according to the frequencies of each in the population $\mathcal{A}_1$ (and handling $\text{Resume}_2$ alignments analogously).} Then it follows from the assumption that $\tau_{\text{SC}}$ and $\tau_{\textsf{audit}}$ are causally consistent that audit studies successfully test the causal effect of race on interview outcomes, in the following sense: the difference in interview status that results from changing $\text{Race}_1$ to $\text{Race}_2$ is the same as the difference that results from changing $\mathcal{A}_1$ to $\mathcal{A}_2$, which is in turn the same as changing $\text{Resume}_1$ and $\text{Resume}_2$. The necessary assumptions of causal consistency are licensed by the social constructionist theory of race that defined $\tau_{SC}$, and by construction of audit studies, which omit all attributes not on the resume via $\tau_{\text{audit}}$ (Figure~\ref{figure:abstraction model}).

\begin{figure}[H]
    \centering
    \begin{tikzpicture}[->, >=stealth, shorten >=1pt, auto, node distance=2cm, thick]
        % Race
        \node[circle, draw, minimum size=5em] (Race) {\textsf{Race}};

        % Resume
        \node[rectangle, dashed, draw, minimum size=5em, right = 50mm of Race] (Resume) {Resume};

        \coordinate (Mid) at ($(Race)!0.5!(Resume)$);
        
         % All attributes
        \node[rectangle, dashed, draw, minimum size=5em, below=25mm of Mid] (features) {All attributes};

        \node[circle,draw,minimum size=5em, right=10mm of Race] (RaceInterview) {\textsf{Interview}};
        \path (Race) edge[->] node {} (RaceInterview);
        
        \node[circle,draw,minimum size=5em, right=10mm of Resume] (ResumeInterview) {\textsf{Interview}};
        \path (Resume) edge[->] node {} (ResumeInterview);
        
        \node[circle,draw,minimum size=5em, right=10mm of features] (featuresInterview) {\textsf{Interview}};
        \path (features) edge[->] node {} (featuresInterview);

        % Alignments
        \path (features) edge[dotted] node[midway,left] {$\tau_{\text{SC}}$} (Race);
        
        \path (features) edge[dotted] node[midway,right] {$\tau_{\text{\text{audit}}}$} (Resume);
        
        \path (featuresInterview) edge[dotted] node[midway,right] {$\tau_{\text{\text{audit}}}$} (ResumeInterview);
        \path (featuresInterview) edge[dotted] node[midway,left] {$\tau_{\text{SC}}$} (RaceInterview);

    \end{tikzpicture}

    \captionsetup{font=small}
    \caption{The abstraction model of audit studies. 
    \new{The alignment $\tau_{\text{SC}}$ is posited by social constructionist theories of race: \textsf{Race} is an abstraction of various attributes and relations an individual stands in, and the abstract causal effect of such a notion of race on interview outcomes is the ground truth target of the audit studies assessing discrimination on the basis of race. (The causal effect of other features on interview outcome can be abstracted away.) Audit studies aim to estimate the causal effect of race on interview outcome by generating resumes that differ only in name, and measuring the difference in interview outcome. That is, they construct a different abstraction of the individual's attributes, namely $\tau_{\text{audit}}$, and the assumption is that toggling only the name in the $\tau_{\text{audit}}$-abstraction is equivalent to toggling race in the $\tau_{\text{SC}}$ abstraction, i.e.\ that these two abstractions are causally consistent. From the social-constructivist perspective the challenge is whether the highly restricted intervention on \textsf{Resume} (toggling name while holding all other resume features fixed) indeed is causally consistent under a socially-constructed account of race.  
    %and the alignment $\tau_{\text{audit}}$ is used by audit studies. Because these alignments are causally consistent, the effect of all attributes on interview status is the same as the effect of resume on interview status, and indeed the same as the effect of race on interview status---at least, when we restrict attention to interventions on constituents of race that change the value of the race variable. Note that in the high-level model, race causes interview status not in the sense that whether a person is interviewed depends deterministically on their race, but rather in the sense that interventions on race change the likelihood with which someone receives an interview (e.g.) to the base rate for that race.
    %
    }
    }
    \label{figure:abstraction model}
\end{figure}

 Formally, audit studies purport to test the effect of race on interview status via the following equation:%\footnote{When the variable $V$ is easily understood from context, we will simply write $\text{do}(v)$ for an intervention. For example, $\text{do}(\textsf{Race}_1)$ sets the $\mathsf{Race}$ variable to value $\mathsf{Race}_1$.}
\begin{align*}
    &\mathbb{P}[\textsf{Interview Callback}\,|\, \text{do}(\text{Race}_1)] - \mathbb{P}[\textsf{Interview Callback}\,|\, \text{do}(\text{Race}_2)] \\
    = \quad &\mathbb{P}[\textsf{Interview Callback}\,|\, \text{do}(\text{Resume}_1)] - \mathbb{P}[\textsf{Interview Callback}\,|\, \text{do}(\text{Resume}_2)].
\end{align*}

Admittedly, this picture is an over-simplification. There are multiple interventions on all of an individual's attributes which could correspond to a change from $\text{Race}_1$ to $\text{Race}_2$, so there is significant under-determination and ambiguity in finding a lower-level intervention to test the effect of this change. But this ambiguity is a common feature of causal inference \citep{spirtes04}: in saying that we changed a stimulus for the bird from red to blue, it is ambiguous whether we changed it from crimson to cyan, or from scarlet to turquoise. Causal consistency guarantees that all such changes produce the same pattern of changes in outcome. Thus, to the extent that we have causal consistency in a diagram like Figure~\ref{figure:abstraction model}, there is a clear and principled sense in which audit studies can be modeled as testing the effect of interventions on race. More generally, one can model claims about causal discrimination as claims about the effects of race, understood as an abstraction of lower-level, in-principle manipulable attributes.\footnote{Interventions on race are practically impossible, since to change someone's race would arguably make them a different person. (One might imagine intervening on a fetus's genes to change its phenotype and assigning the fetus to a mother of a race associated with that phenotype (\cite{marcellesi2013race}, p. 655), but this is clearly not the kind of intervention that social scientists, U.S. courts, and  computer scientists have in mind when discussing causal discrimination (\cite{weinberger22}, p. 1269).) This impossibility can seem to threaten the intelligibility of modeling race as a cause, and thus of the notion of causal discrimination (cf. \citealt{holland1986statistics}, p. 946; \citealt{greiner2011causal}, pp. 1-2; \citealt{glymour2014race}; \citealt{sen2016race}, p. 504). However, the impossibility of intervening on race is not an insurmountable problem \citep{pearl2018does}. As \cite{weinberger22} observes (p. 1268), while it is impossible (due to regulations) to intervene on people's smoking habits, we can still estimate causal effects of smoking: upon observing different health outcomes for relevantly similar populations that differ in their smoking habits, we can often reasonably infer that smoking caused these outcomes. We can even make this inference when a person would have been very different, had they not been a smoker. Similarly, upon observing different outcomes for relevantly similar individuals of different races, we can often reasonably infer that race caused this difference in outcomes (cf. \citealt{marcellesi2013race}, p. 656)---though of course one must specify what counts as relevant similarity, which we propose to do using the theory of causal abstraction.}

It is sometimes objected that structural causal models cannot accommodate constitutive relations, because constitutive relations lack the directionality characteristic of causal relationships \citep[p.5]{hu2020sex}. As the above diagram illustrates, this objection can be overcome by appeal to abstraction. One might argue that the specified alignments do not commute with the intervention performed by audit studies, and thus that the crucial assumption of causal consistency fails to hold even approximately; we return to this worry below in Section~\ref{subsec:distinguish intervention identification challenges}. However, we show in the next section that given this assumption, the abstraction framework addresses the modularity problem.

\section{The modularity problem}\label{sec:modularity}

In this section, we explain the modularity requirement and the modularity problem for causal models of discrimination. We then explain how the abstraction framework for modeling discrimination addresses this problem and compare it to some alternatives.

\subsection{The modularity requirement}\label{subsec:modularity:introducing}

In structural causal models, $X$ causes $Y$ when changes in $X$ lead to changes in $Y$. For instance, to determine if there is a causal relationship between object color and pecking behavior, we could try intervening on each independently, and seeing which intervention leads to a change in the other variable. In this case, we would observe that changing the color of an object (for instance, by painting it) leads to a change in bird’s pecking behavior, but that changing the bird’s pecking behavior (for instance, by bribing the bird with a treat) does not change the object’s color. We thus conclude that the object's color causes pecking behavior and not the other way around.

To allow for these kinds of interventions on individual variables, structural causal models must satisfy a modularity requirement (see, e.g., \citealt{Pearl2000-PEAC-2}, p. 63; \citealt{woodward2005making},p.327; \cite{peters2017elements}, pp.17-20, point 1). A variable in a causal model is \textit{modular} when it can be intervened upon independently from intervening on the others, and a causal model is \textit{modular} when all its variables are modular. % TO DO: Add citation

For example, a causal model which implies that object color causes bird pecking behavior is modular because it is possible to conceive of interventions on each of the two variables that do not act on the other. If we paint an object blue, we perform an intervention that directly changes just that object’s color, not whether the bird is wont to peck at blue objects.\footnote{Of course, since variables often causally influence other variables, modularity allows for an intervention on some variable to affect other variables through \textit{causal} relations. Interventions on the color of the object plainly affect the bird’s actual pecking causally, but color and pecking behavior are still modular. Instead, when a variable $X$ is modular with respect to $Y$, this means that interventions on $X$ can be conceptually separated from interventions on $Y$. While this distinction between causal and non-causal relations is an intuitive one, \cite{janzing2022phenomenologicalcausality} discuss a number of puzzles about how to make it fully precise.} As a result, any effects of this intervention demonstrate the causal relationship between object color and bird pecking behavior.

Without modularity, the causal effects of an intervention on a variable are hopelessly conflated with the intervention itself. Consider, for instance, a causal model claiming that the coarse color of an object causes its particular fine shade. This model is non-modular because it is conceptually impossible to manipulate the color of an object separately from intervening on its particular shade (and often, vice versa). How could we change the color of an object from red to blue, without changing it from some shade of red to some shade of blue? Since actions on fine shades and coarse colors and are muddled together, we cannot isolate a causal effect of the intervention.

\subsection{Introducing the modularity problem}\label{sec:modularity:problem}

Return to the resume audit study, in which experimenters aim to establish a causal effect of race on interview decisions by performing manipulations on race-coded names. To support reasoning about whether race (via manipulations on names) causes interview callbacks, we should be able to represent the audit study using a causal model with variables corresponding to race, name, and interview callback status. This causal model must include names and interview callbacks to represent the causal effect of name interventions on interview callbacks performed by the audit study, and it must include race to explain (by reference to some property of the causal model) how this name manipulation reveals an effect of race.

The modularity problem, developed by Lily Hu and Issa Kohler-Hausmann \citep{kohlerhausmann2018eddie,hu2020sex}, is that attributes like race, gender, and religion cannot be placed in a single causal model with the attributes from which they are socially constructed, because their constitutive relations violate the modularity requirement. In the case of audit studies, the problem is that race, names, and interview callback status cannot be placed in a single structural causal model without violating the modularity assumption crucial to causal modeling and inference. The worry is that these three variables interact more like colors and their more specific shades than like bird pecking behavior and object colors; models including all three variables inevitably end up non-modular.

The problem is easiest to see if we accept that race is a social construction, and thus stands in \textit{constitutive} rather than \textit{causal} relations to many of its correlates. For then certain names constitute part of the social meaning of race, and thus race itself. For instance, Black-sounding names like `Jamal' form part of the social meaning of what it is to be Black. As a result, an intervention setting a name of an applicant on a resume to Jamal cannot, in principle, be separated from an intervention to their race, and vice versa.  Thus, placing race and names in a causal model together --- as formalizations of audit studies do, to justify using an intervention on names to reveal an effect of race --- requires making a false modularity assumption that names do not partially constitute race. This problem can arise for any attributes that appear on a resume, like education or employment history, whenever they partly constitute race.

\subsection{Abstraction as a solution to the modularity problem}

The abstraction model of discrimination addresses the modularity problem. Because socially constructed attributes and their constituents are placed in \textit{separate causal models}, and modularity need only hold within each model taken separately, there is no non-modularity within either the high-level model containing race or the low-level model containing its constituents. Audit studies can be understood as testing the causal effects of race, by testing the effects of its constituents, as defined by a causally consistent alignment. The relationship specified by this alignment is non-modular, in the sense that changes to names correspond to changes in race, but this is a benign (indeed intended and necessary) feature of causally consistent alignments, not an obstacle to causal modeling: when one attribute is simply a higher-level abstraction of other, lower-level attributes, one should expect changes at either level to correspond to changes at the other.\footnote{The abstraction model does leave the outcome variable---in this case, interview status---out of the alignment. It may be objected that interview status is constitutive of race, and thus that the modularity problem persists. However, in the resume audit study, the outcome variable does not represent the interview status of the applicant for all jobs, but is specific to the decision audited in the study. Since that decision has not happened yet at the point of the audit, it cannot yet be constitutive of race, and thus poses no modularity problem. We agree that on a functional understanding of race as a social kind that sustains social stratification, the causal capacity of race to reinforce the hierarchical social positions that constitute it may be thought of as part of what defines race. But this capacity is better represented by a causal relation between race and future social positions, rather than by inclusion of those social positions in the constitutive basis from which race is abstracted.}

By dissolving the modularity problem, the abstraction model immediately dissipates the concern that the audit studies fail to demonstrate any intelligible causal effect of race at all. This in turn shows that the modularity requirement does not in itself force causal models to posit an especially implausible metaphysical boundary between race and the context in which it is socially embedded \citep{kohlerhausmann2018eddie, hu2020sex, hu2024}. Instead, this relationship is modeled within the framework, via the causally consistent alignment $\tau_{\text{SC}}$ that specifies how race is socially constructed. This alignment may include all the attributes that are included in a particular social constructionist theory of race---for instance, names, education history, and employment history.

% Of course, the use of abstraction does not obviate the need to make substantive normative judgments in causal modeling about how race and its effects are abstracted from low-level attributes. Nor does it tell us, once we have an abstraction, how to choose the relevant counterfactual contrasts in testing for the causal effects of race or other social kinds. These are further modeling choices, which are subject to empirical and normative challenges. We discuss these issues in Section~\ref{sec:further challenges}.

\subsection{Comparison to other approaches}

Our proposal formalizes and further develops Sen and Wasow's~\citeyearpar{sen2016race} suggestion to treat race as a ``bundle of sticks,'' such that manipulations on race are manipulations on some of the sticks. \cite{kohlerhausmann2018eddie} raises a modularity-like worry for this proposal: causal manipulations on constituents of race might alter the constitutive organization and causal profile of race, making the ``constitutive graph'' unstable. For instance, an audit study on names might flood a city's job market with so many resumes for Jamal that it changes perceptions of how common the name Jamal is among Black people, in turn changing how names constitute the social kind of race in that city.\footnote{Just like non-modularity, this distribution-dependence of the relations between variables threatens the intelligibility of causal models. The so-called principle of independent mechanisms that motivates modularity can be interpreted to imply that for all variables $V$ in a causal model, the mechanism $f_V$ by which it depends on its causal parents and exogenous noise does not contain information about the distribution of any other variable $X$ (see \cite{janzingCausalInferenceUsing2008}; \cite{peters2017elements} pp.17-21, point 2 and pp.58-62). This requirement is sometimes also referred to as ``modularity'' but is distinct from the notion of modularity we presented  in \S\ref{subsec:modularity:introducing}.} 

Fortunately, these kinds of distribution-dependencies are compatible with causal abstraction. For example, one can represent virtual memory in a computer as an abstraction of physical memory in that computer, even though the alignment between virtual and physical memory itself depends on the distribution of physical memory.\footnote{Thanks to Atticus Geiger for the example.}
Analogously, the alignment from lower-level attributes to race may itself depend on the distribution of those lower-level attributes. For example, the ability of names like ``Jamal'' to count as race-coded plausibly depends on how this name is perceived by recruiters, and thus on its distribution in the population. When these distributional dependencies are idealized away (as they may be in small-scale experimental manipulations like those of most audit studies, as opposed to sweeping interventions such as policy changes), abstraction can provide a formalism to state assumptions about their absence, thereby exposing them to empirical challenges.

Concerns about race's manipulability led \cite{greiner2011causal} to propose that we use the perception of race, rather than race itself, as a treatment in causal studies of race. But replacing race by its perception simply pushes the modularity problem one step back: if race is socially constructed, we should expect interventions on the perception of race to be bound up with interventions on the perception of names, education, and so on.\footnote{As \cite{weinberger22} observes, it is unclear that when a victim is harmed based on a mistaken perception of their race, we should count this as wrongful racial discrimination. Whether a person actually possesses the race in question surely informs the way in which the act is wrong, and the degree to which it is wrong. This is another worry for replacing race with its perception.} 

\cite{weinberger22} suggests, more generally, that we can model audit studies as manipulating a signal of race that is causally downstream of race. Audit studies vary names ``as if'' they were varying race, thereby revealing an effect of race along a particular causal path.\footnote{\cite{vanderweele2014causal} propose a model of race as a cause which is very similar to Weinberger's, in which socioeconomic status, physical phenotype, parental physical phenotype, genetic background, cultural context, family, and neighborhood function as signals of race (in Weinberger's sense). This proposal faces the same worries we raise for Weinberger's view. We focus on the latter because it is framed as a response to the modularity problem.} To include race and names in the same causal model without violating modularity, Weinberger argues that it does not follow from the assumption that race is socially constructed that race is constituted by its signals (e.g., names). If race is not constituted by its signals, then it is possible to manipulate those signals separately from race.

\new{If this is going to work, the appeal to signals will need to model \textit{every} would-be source of modularity as a signal (rather than a constituent) of race. Not only names, but also university, socioeconomic status, relations of subordination, and so on will need to be called signals (not constituents) of race. When this is done, we seem to end up with a ``thin constructionist'' view of race, which says race is constituted by just a few superficial features of people's appearances and ancestry \citep{mallonRaceNormativeNot2006}. We worry that this just rejects the thicker view of social constructionism that motivates the problem we wanted to solve (\citeauthor{hu2024}, 2024, §4; \citeauthor{huWhatPerceivedWhen2024}, 2024, pp. 249-51). The signal-manipulation approach will also struggle to plausibly model cases where racial categorization comes apart from appearance and ancestry (\citealt{mills1998blackness}, cf. Section~\ref{sec:abstraction:social_construction}). By contrast, the abstraction model does not require us to reject the social constructionist views that motivate the modularity problem: it allows that race may be partly constituted by signals like names.}\footnote{\new{Weinberger proposes that in many narrow experimental contexts, signals like names send sufficiently ``isolated'' signals of individuals’ race, such that modularity can be adopted as a limited modeling assumption \citep[p. 1282]{weinberger22}. But we worry that this stipulation can seem overly optimistic, given the empirical evidence (summarized by Hu~\citeyear{huWhatPerceivedWhen2024}) that racial perception, even if induced just by a name, starts a cascade of racialized interpretation of other attributes that at first blush do not convey racial meaning, like what university applicants attended, their hometown, or their work experience.}}

\new{This illustrates a more general point. We can try to model interactions between race and its signals (and among its signals), for example by adding interaction terms to the model. But it only makes sense to consider race and its signals in a single-level model if there is no modularity problem, because non-modularity makes interventions in such a model ill-defined. So modeling interactions in this way just assumes away the modularity problem. By contrast, abstraction directly addresses the problem, by providing  a framework for explicitly modeling the non-modular relationships between race and its signals. In Section~\ref{sec:modularity:skepticism}, we argue that abstraction remains a fruitful framework, even if we just deny the modularity problem and reject the view of social construction that motivates it.}

\subsection{Skepticism about social constructionism}\label{sec:modularity:skepticism}

\new{Needless to say, not everyone is a thick social constructionist---one might reject thick social constructionism and opt for thin social constructionism (Section \ref{sec:abstraction:social_construction}), or reject social constructionism altogether.\footnote{\new{For example, on Spencer's~\citeyearpar{Spencer2019} view, races are possibly overlapping genetic clusters of people, called \textit{human continental populations}. On this view, race is not socially constructed, but rather biologically real, because scientific expert usage determines the constitution of race \citep[pp. 82-3]{Spencer2019}, and races (understood as human continental populations) are epistemically useful and justified entities in a well-ordered research program in biology, namely population genetics \cite[pp. 95-6]{Spencer2019}. We note that Spencer's~\citeyearpar{Spencer2019RadicalSolution} later view is pluralist, countenancing several different notions of race.}} While objections to social constructionism do not provide a way of addressing the modularity problem on its own terms, they raise the question of whether the abstraction framework has anything to offer those who reject social constructionism. Our primary aim in this paper is to show that abstraction preserves a precise causal account of discrimination, by allowing us to explicitly model assumptions about social constructions, which might then be challenged, debated, or verified. But in closing this section, we want to underscore that the framework of causal abstraction does not itself imply social constructionism, and that it can be given independent motivation.}

\new{Tests for causal discrimination are often \textit{path-specific}. State-of-the-art causal fairness criteria specify that only effects along certain pathways from race-related variables to the outcome of interest should count as discriminatory (\citeauthor{zhang2018fairness}, \citeyear{zhang2018fairness}, p. 2; \citeauthor{nilforoshan2022causal}, \citeyear{nilforoshan2022causal}, p.~4;
\citeauthor{weinberger22}, \citeyear{weinberger22}, pp.~1276--7). Suppose we have figured out which paths are relevant, and the effects of some particular variables $V_1,\dots,V_n$ on the outcome of interest amount to discrimination. For example, $V_1$ could be a person's appearance, and $V_2$ could be their ancestry, and so on. Now, the overall measure of discrimination will be some function $f$ of the effects of these variables---a weighted average, say.}

\new{If it really makes sense to talk about $f$ as capturing the cause of discrimination, we should expect that $f$ satisfies some notion of approximate causal consistency. But this is just the abstraction framework we are proposing. We can view $f$ as a causally consistent alignment between low-level variables $V_1,\dots,V_n$ and a high-level variable $V$. This variable $V$ could represent race, as we have not said that $V_1,\dots,V_n$ are effects (as opposed to constituents) of race. But if we are not social constructionists, this variable $V$ does not have to itself represent race. It could represent (e.g.) racialization, or race as it operates in discrimination. The abstraction framework captures a fundamental idea, namely that the measure of  discrimination should incorporate different factors in a causally consistent way. The variable representing the aggregate of these factors can be interpreted by the social constructionist as the protected attribute itself, but this interpretation is not required by the formal framework.}

\section{The intervention identification problem}\label{sec:further challenges}

We have argued that the framework of causal abstraction provides a natural way of capturing the constitutive relations between race and the attributes from which it is socially constructed (Section~\ref{sec:abstraction}) and that this framework addresses the modularity problem raised for causal models of discrimination (Section~\ref{sec:modularity}).

However, there remains the \textit{intervention identification problem} of explaining when and why interventions reveal causal effects of a social kind normatively relevant to discrimination. To illustrate, consider the following examples:

\begin{itemize}

 \item 
    \textit{U.S. law.} In \textit{Bostock vs. Clayton}, the plaintiff, a gay male employee, was fired for his sexual orientation. The majority decided that this amounted to sexual discrimination, since had the male employee been a female employee attracted to men, she would not have been fired. The dissenters replied that had the male employee been a female employee with same-sex attraction, she would have still been fired  \citep{dembroff2020taylor,kohler-hausmannSupremeConfusionCausality2022}. Should we keep sexual orientation fixed, when testing the effect of sex on firing decisions? 

    \item 
    \textit{Audit studies.} \new{Suppose we designed an in-person audit study with two actors, one a man and the other a woman, to test the effect of gender on interview outcomes. This study holds everything but applicants' gender fixed: the actors have similar resumes, timeliness, and answers to interview questions, and both actors wear makeup and skirts. Proponents of this study might claim that it successfully tests gender discrimination, because all variables except for gender are held fixed. Dissenters would reply that gender is partly constituted by norms regarding clothing and makeup, such that these factors must be varied between the actors in order to test the effect of gender} (\citealt{huNormativeFactsCausal2024} p. 8; cf. \citealt{kohlerhausmann2018eddie}, p. 1216). Should we keep clothing and makeup fixed when intervening on gender to test its effect on hiring decisions? 
\end{itemize}

\noindent In each case, we face an intervention identification problem: when and why should attributes be varied with a protected attribute, such that worse treatment on their basis constitutes discrimination? This is not a purely metaphysical question, but also a normative one; it arises because there is no non-normative, purely metaphysical boundary between social kinds like race or gender and the context in which they are socially embedded.

In this section, we use the abstraction framework to distinguish a number of further worries for causal models of discrimination, showing that these are ultimately objections to particular models  of discrimination, rather than challenges to causal models of discrimination as such (Section~\ref{subsec:distinguish intervention identification challenges}). We then discuss how these debates about modeling assumptions bear on experimental design (Section~\ref{subsec:experimental design}).

\subsection{Challenges of intervention identification}\label{subsec:distinguish intervention identification challenges}

We can now use the framework of causal abstraction to distinguish two challenges to the claim that by intervening on names, audit studies reveal an effect that is normatively significant and a sign of discrimination.

Consider, first, a set of challenges associated with defining the alignment and high-level model. An initial question is how to decide what features belong in the constitutive basis of race. \new{But even once these features are chosen, interventions on race will be ambiguous between lower-level interventions on its many constituents (cf. Tolbert~\citeyear[pp. 1103-5]{tolbertcausal}), which may have different effects. Indeed, even if the changes to the names on resumes performed in audit studies correspond, via the relevant alignments, to changes in race, there are many such changes. For example, changing university from ``Harvard'' to ``Howard'' would typically also change the race most likely associated with a given resume, and may well have a different effect on interview status than changing ``Greg'' to ``Jamal'' (Figure~\ref{fig:ambiguity}).}

\begin{figure}[H] \centering
\subfigure [Coarse-grained] {
\begin{tikzpicture}
% first row
      \node at (0,0) [fill=teal!55] {};
      \node at (.25,0) [fill=teal!55] {};
      \node at (.5,0) [fill=teal!55] {};
      \node at (0,.25) [fill=teal!55] {};
      \node at (.25,.25) [fill=teal!55] {};
      \node at (.5,.25) [fill=teal!55] {};
      \node at (0,.5) [fill=teal!55] {};
      \node at (.25,.5) [fill=teal!55] {};
      \node at (.5,.5) [fill=teal!55] {};
      
      \node at (.75,0) [fill=teal!88] {};
      \node at (1,0) [fill=teal!88] {};
      \node at (1.25,0) [fill=teal!88] {};
      \node at (.75,.25) [fill=teal!88] {};
      \node at (1,.25) [fill=teal!88] {};
      \node at (1.25,.25) [fill=teal!88] {};
      \node at (.75,.5) [fill=teal!88] {};
      \node at (1,.5) [fill=teal!88] {};
      \node at (1.25,.5) [fill=teal!88] {};
      
      \node at (1.5,0) [fill=teal!22] {};
      \node at (1.75,0) [fill=teal!22] {};
      \node at (2,0) [fill=teal!22] {};
      \node at (1.5,.25) [fill=teal!22] {};
      \node at (1.75,.25) [fill=teal!22]{};
      \node at (2,.25) [fill=teal!22] {};
      \node at (1.5,.5) [fill=teal!22]{};
      \node at (1.75,.5) [fill=teal!22] {};
      \node at (2,.5) [fill=teal!22] {};
      
      \node at (2.25,0) [fill=teal!33] {};
      \node at (2.5,0) [fill=teal!33] {};
      \node at (2.75,0) [fill=teal!33] {};
      \node at (2.25,.25) [fill=teal!33] {};
      \node at (2.5,.25) [fill=teal!33] {};
      \node at (2.75,.25) [fill=teal!33] {};
      \node at (2.25,.5) [fill=teal!33] {};
      \node at (2.5,.5) [fill=teal!33] {};
      \node at (2.75,.5) [fill=teal!33] {};
      
 % second row:
 
      \node at (0,.75) [fill=teal!77] {};
      \node at (.25,.75) [fill=teal!77] {};
      \node at (.5,.75) [fill=teal!77] {};
      \node at (0,1) [fill=teal!77] {};
      \node at (.25,1) [fill=teal!77] {};
      \node at (.5,1) [fill=teal!77] {};
      \node at (0,1.25) [fill=teal!77] {};
      \node at (.25,1.25) [fill=teal!77] {};
      \node at (.5,1.25) [fill=teal!77] {};

      \node at (.75,.75) [fill=teal!33] {};
      \node at (1,.75) [fill=teal!33] {};
      \node at (1.25,.75) [fill=teal!33] {};
      \node at (.75,1) [fill=teal!33] {};
      \node at (1,1) [fill=teal!33] {};
      \node at (1.25,1) [fill=teal!33] {};
      \node at (.75,1.25) [fill=teal!33] {};
      \node at (1,1.25) [fill=teal!33] {};
      \node at (1.25,1.25) [fill=teal!33] {};

      \node at (.99,.99) [rectangle,draw=pink,very thick,minimum width=7.25mm,minimum height=7.25mm] {};
      
      \node at (1.5,.75) [fill=teal!66] {};
      \node at (1.75,.75) [fill=teal!66] {};
      \node at (2,.75) [fill=teal!66] {};
      \node at (1.5,1) [fill=teal!66] {};
      \node at (1.75,1) [fill=teal!66] {};
      \node at (2,1) [fill=teal!66] {};
      \node at (1.5,1.25) [fill=teal!66] {};
      \node at (1.75,1.25) [fill=teal!66] {};
      \node at (2,1.25) [fill=teal!66] {};
      
      \node at (2.25,.75) [fill=teal!55] {};
      \node at (2.5,.75) [fill=teal!55] {};
      \node at (2.75,.75) [fill=teal!55] {};
      \node at (2.25,1) [fill=teal!55] {};
      \node at (2.5,1) [fill=teal!55] {};
      \node at (2.75,1) [fill=teal!55] {};
      \node at (2.25,1.25) [fill=teal!55] {};
      \node at (2.5,1.25) [fill=teal!55] {};
      \node at (2.75,1.25) [fill=teal!55] {};
      
% third row:

      \node at (0,1.5) [fill=teal!22] {};
      \node at (.25,1.5) [fill=teal!22] {};
      \node at (.5,1.5) [fill=teal!22] {};
      \node at (0,1.75) [fill=teal!22] {};
      \node at (.25,1.75) [fill=teal!22] {};
      \node at (.5,1.75) [fill=teal!22] {};
      \node at (0,2) [fill=teal!22] {};
      \node at (.25,2) [fill=teal!22] {};
      \node at (.5,2) [fill=teal!22] {};
      
      \node at (.75,1.5) [fill=teal!55] {};
      \node at (1,1.5) [fill=teal!55] {};
      \node at (1.25,1.5) [fill=teal!55] {};
      \node at (.75,1.75) [fill=teal!55] {};
      \node at (1,1.75) [fill=teal!55] {};
      \node at (1.25,1.75) [fill=teal!55] {};
      \node at (.75,2)[fill=teal!55] {};
      \node at (1,2) [fill=teal!55] {};
      \node at (1.25,2) [fill=teal!55] {};
      
      \node at (1.5,1.5) [fill=teal!77] {};
      \node at (1.75,1.5) [fill=teal!77] {};
      \node at (2,1.5) [fill=teal!77] {};
      \node at (1.5,1.75) [fill=teal!77] {};
      \node at (1.75,1.75) [fill=teal!77] {};
      \node at (2,1.75) [fill=teal!77] {};
      \node at (1.5,2) [fill=teal!77] {};
      \node at (1.75,2) [fill=teal!77] {};
      \node at (2,2) [fill=teal!77] {};

      \node at (2.25,1.5) [fill=teal] {};
      \node at (2.5,1.5) [fill=teal] {};
      \node at (2.75,1.5) [fill=teal] {};
      \node at (2.25,1.75) [fill=teal] {};
      \node at (2.5,1.75) [fill=teal] {};
      \node at (2.75,1.75) [fill=teal] {};
      \node at (2.25,2) [fill=teal] {};
      \node at (2.5,2) [fill=teal] {};
      \node at (2.75,2) [fill=teal] {};
      
% fourth row:
      \node at (0,2.25) [fill=teal!77] {};
      \node at (.25,2.25) [fill=teal!77] {};
      \node at (.5,2.25) [fill=teal!77] {};
      \node at (0,2.5) [fill=teal!77] {};
      \node at (.25,2.5) [fill=teal!77] {};
      \node at (.5,2.5) [fill=teal!77] {};
      \node at (0,2.75) [fill=teal!77] {};
      \node at (.25,2.75) [fill=teal!77] {};
      \node at (.5,2.75) [fill=teal!77] {};
      
      \node at (.75,2.25) [fill=teal!44] {};
      \node at (1,2.25) [fill=teal!44] {};
      \node at (1.25,2.25) [fill=teal!44] {};
      \node at (.75,2.5) [fill=teal!44] {};
      \node at (1,2.5) [fill=teal!44] {};
      \node at (1.25,2.5) [fill=teal!44] {};
      \node at (.75,2.75) [fill=teal!44] {};
      \node at (1,2.75) [fill=teal!44] {};
      \node at (1.25,2.75) [fill=teal!44] {};
      
      \node at (1.5,2.25) [fill=teal!33] {};
      \node at (1.75,2.25) [fill=teal!33] {};
      \node at (2,2.25) [fill=teal!33] {};
      \node at (1.5,2.5) [fill=teal!33] {};
      \node at (1.75,2.5) [fill=teal!33] {};
      \node at (2,2.5) [fill=teal!33] {};
      \node at (1.5,2.75) [fill=teal!33] {};
      \node at (1.75,2.75) [fill=teal!33] {};
      \node at (2,2.75) [fill=teal!33] {};
      
      \node at (2.25,2.25) [fill=teal!55] {};
      \node at (2.5,2.25) [fill=teal!55] {};
      \node at (2.75,2.25) [fill=teal!55] {};
      \node at (2.25,2.5) [fill=teal!55] {};
      \node at (2.5,2.5) [fill=teal!55] {};
      \node at (2.75,2.5) [fill=teal!55] {};
      \node at (2.25,2.75) [fill=teal!55] {};
      \node at (2.5,2.75) [fill=teal!55] {};
      \node at (2.75,2.75) [fill=teal!55] {};
      
 % Counterfactuals:

       \node at (.99,.99) [rectangle,draw=red!70,very thick,minimum width=7.25mm,minimum height=7.25mm] {};

        \node at (1.74,1.74) [rectangle,draw=red!70,very thick,minimum width=7.25mm,minimum height=7.25mm] {};
 
     \node (p1) at (1.15,.9) {};
     \node (p2) at (1.9,1.75) {}; 
     \path (p1) edge[->,very thick,bend left,color=black] (p2); 
     \node at (1.9,1.8) {\large{}\textcolor{white}{\textbf{?}}};
      
\end{tikzpicture}} \hspace{1.25in} 
\subfigure [Fine-grained] {
\begin{tikzpicture}
% first row:
      \node at (0,0) [fill=teal,draw=gray!50] {};
      \node at (.25,0) [draw=gray!50] {};
      \node at (.5,0) [fill=teal,draw=gray!50] {};
      \node at (0,.25) [fill=teal,draw=gray!50] {};
      \node at (.25,.25) [draw=gray!50] {};
      \node at (.5,.25) [draw=gray!50] {};
      \node at (0,.5) [draw=gray!50] {};
      \node at (.25,.5) [fill=teal,draw=gray!50] {};
      \node at (.5,.5) [fill=teal,draw=gray!50] {};
      
      \node at (.75,0) [fill=teal,draw=gray!50] {};
      \node at (1,0) [fill=teal,draw=gray!50] {};
      \node at (1.25,0) [fill=teal,draw=gray!50] {};
      \node at (.75,.25) [fill=teal,draw=gray!50] {};
      \node at (1,.25) [draw=gray!50] {};
      \node at (1.25,.25) [fill=teal,draw=gray!50] {};
      \node at (.75,.5) [fill=teal,draw=gray!50] {};
      \node at (1,.5) [fill=teal,draw=gray!50] {};
      \node at (1.25,.5) [fill=teal,draw=gray!50] {};
      
      \node at (1.5,0) [draw=gray!50] {};
      \node at (1.75,0) [fill=teal,draw=gray!50] {};
      \node at (2,0) [draw=gray!50] {};
      \node at (1.5,.25) [draw=gray!50] {};
      \node at (1.75,.25) [draw=gray!50] {};
      \node at (2,.25) [fill=teal,draw=gray!50] {};
      \node at (1.5,.5) [draw=gray!50] {};
      \node at (1.75,.5) [draw=gray!50] {};
      \node at (2,.5) [draw=gray!50] {};
      
      \node at (2.25,0) [draw=gray!50] {};
      \node at (2.5,0) [fill=teal,draw=gray!50] {};
      \node at (2.75,0) [fill=teal,draw=gray!50] {};
      \node at (2.25,.25) [draw=gray!50] {};
      \node at (2.5,.25) [draw=gray!50] {};
      \node at (2.75,.25) [draw=gray!50] {};
      \node at (2.25,.5) [draw=gray!50] {};
      \node at (2.5,.5) [fill=teal,draw=gray!50] {};
      \node at (2.75,.5) [draw=gray!50] {};
      
 % second row:
      \node at (0,.75) [fill=teal,draw=gray!50] {};
      \node at (.25,.75) [fill=teal,draw=gray!50] {};
      \node at (.5,.75) [draw=gray!50] {};
      \node at (0,1) [fill=teal,draw=gray!50] {};
      \node at (.25,1) [fill=teal,draw=gray!50] {};
      \node at (.5,1) [draw=gray!50] {};
      \node at (0,1.25) [fill=teal,draw=gray!50] {};
      \node at (.25,1.25) [fill=teal,draw=gray!50] {};
      \node at (.5,1.25) [fill=teal,draw=gray!50] {};
      
      \node at (.75,.75) [draw=gray!50] {};
      \node at (1,.75) [fill=teal,draw=gray!50] {};
      \node at (1.25,.75) [fill=teal,draw=gray!50] {};
      \node at (.75,1) [draw=gray!50] {};
      \node at (1,1) [fill=teal,draw=gray!50] {};
      \node at (1.25,1) [draw=gray!50] {};
      \node at (.75,1.25) [fill=teal,draw=gray!50] {};
      \node at (1,1.25) [draw=gray!50] {};
      \node at (1.25,1.25) [draw=gray!50] {};
      
      \node at (1.5,.75) [draw=gray!50] {};
      \node at (1.75,.75) [draw=gray!50] {};
      \node at (2,.75) [draw=gray!50] {};
      \node at (1.5,1) [draw=gray!50] {};
      \node at (1.75,1) [fill=teal,draw=gray!50] {};
      \node at (2,1) [fill=teal,draw=gray!50] {};
      \node at (1.5,1.25) [draw=gray!50] {};
      \node at (1.75,1.25) [fill=teal,draw=gray!50] {};
      \node at (2,1.25) [draw=gray!50] {};
      
      \node at (2.25,.75) [fill=teal,draw=gray!50] {};
      \node at (2.5,.75) [fill=teal,draw=gray!50] {};
      \node at (2.75,.75) [draw=gray!50] {};
      \node at (2.25,1) [fill=teal,draw=gray!50] {};
      \node at (2.5,1) [draw=gray!50] {};
      \node at (2.75,1) [draw=gray!50] {};
      \node at (2.25,1.25) [fill=teal,draw=gray!50] {};
      \node at (2.5,1.25) [draw=gray!50] {};
      \node at (2.75,1.25) [fill=teal,draw=gray!50] {};
      
% third row:

      \node at (0,1.5) [draw=gray!50] {};
      \node at (.25,1.5) [draw=gray!50] {};
      \node at (.5,1.5) [draw=gray!50] {};
      \node at (0,1.75) [draw=gray!50] {};
      \node at (.25,1.75) [fill=teal,draw=gray!50] {};
      \node at (.5,1.75) [draw=gray!50] {};
      \node at (0,2) [fill=teal,draw=gray!50] {};
      \node at (.25,2) [draw=gray!50] {};
      \node at (.5,2) [draw=gray!50] {};
      
      \node at (.75,1.5) [fill=teal,draw=gray!50] {};
      \node at (1,1.5) [draw=gray!50] {};
      \node at (1.25,1.5) [draw=gray!50] {};
      \node at (.75,1.75) [fill=teal,draw=gray!50] {};
      \node at (1,1.75) [fill=teal,draw=gray!50] {};
      \node at (1.25,1.75) [fill=teal,draw=gray!50] {};
      \node at (.75,2) [draw=gray!50] {};
      \node at (1,2) [draw=gray!50] {};
      \node at (1.25,2) [fill=teal,draw=gray!50] {};
      
      \node at (1.5,1.5) [draw=gray!50] {};
      \node at (1.75,1.5) [fill=teal,draw=gray!50] {};
      \node at (2,1.5) [fill=teal,draw=gray!50] {};
      \node at (1.5,1.75) [fill=teal,draw=gray!50] {};
      \node at (1.75,1.75) [draw=gray!50] {};
      \node at (2,1.75) [fill=teal,draw=gray!50] {};
      \node at (1.5,2) [fill=teal,draw=gray!50] {};
      \node at (1.75,2) [fill=teal,draw=gray!50] {};
      \node at (2,2) [fill=teal,draw=gray!50] {};
      
      \node at (2.25,1.5) [fill=teal,draw=gray!50] {};
      \node at (2.5,1.5) [fill=teal,draw=gray!50] {};
      \node at (2.75,1.5) [fill=teal,draw=gray!50] {};
      \node at (2.25,1.75) [fill=teal,draw=gray!50] {};
      \node at (2.5,1.75) [fill=teal,draw=gray!50] {};
      \node at (2.75,1.75) [fill=teal,draw=gray!50] {};
      \node at (2.25,2) [fill=teal,draw=gray!50] {};
      \node at (2.5,2) [fill=teal,draw=gray!50] {};
      \node at (2.75,2) [fill=teal,draw=gray!50] {};
      
% fourth row:
      \node at (0,2.25) [fill=teal,draw=gray!50] {};
      \node at (.25,2.25) [fill=teal,draw=gray!50] {};
      \node at (.5,2.25) [fill=teal,draw=gray!50] {};
      \node at (0,2.5) [fill=teal,draw=gray!50] {};
      \node at (.25,2.5) [fill=teal,draw=gray!50] {};
      \node at (.5,2.5) [draw=gray!50] {};
      \node at (0,2.75) [fill=teal,draw=gray!50] {};
      \node at (.25,2.75) [draw=gray!50] {};
      \node at (.5,2.75) [fill=teal,draw=gray!50] {};
      
      \node at (.75,2.25) [draw=gray!50] {};
      \node at (1,2.25) [fill=teal,draw=gray!50] {};
      \node at (1.25,2.25) [fill=teal,draw=gray!50] {};
      \node at (.75,2.5) [draw=gray!50] {};
      \node at (1,2.5) [draw=gray!50] {};
      \node at (1.25,2.5) [draw=gray!50] {};
      \node at (.75,2.75) [fill=teal,draw=gray!50] {};
      \node at (1,2.75) [draw=gray!50] {};
      \node at (1.25,2.75) [fill=teal,draw=gray!50] {};
      
      \node at (1.5,2.25) [fill=teal,draw=gray!50] {};
      \node at (1.75,2.25) [fill=teal,draw=gray!50] {};
      \node at (2,2.25) [draw=gray!50] {};
      \node at (1.5,2.5) [fill=teal,draw=gray!50] {};
      \node at (1.75,2.5) [draw=gray!50] {};
      \node at (2,2.5) [draw=gray!50] {};
      \node at (1.5,2.75) [draw=gray!50] {};
      \node at (1.75,2.75) [draw=gray!50] {};
      \node at (2,2.75) [draw=gray!50] {};
      
      \node at (2.25,2.25) [fill=teal,draw=gray!50] {};
      \node at (2.5,2.25) [fill=teal,draw=gray!50] {};
      \node at (2.75,2.25) [draw=gray!50] {};
      \node at (2.25,2.5) [draw=gray!50] {};
      \node at (2.5,2.5) [draw=gray!50] {};
      \node at (2.75,2.5) [draw=gray!50] {};
      \node at (2.25,2.75) [fill=teal,draw=gray!50] {};
      \node at (2.5,2.75) [fill=teal,draw=gray!50] {};
      \node at (2.75,2.75) [fill=teal,draw=gray!50] {};
      
 % Counterfactuals:

             \node at (1,1) [rectangle,draw=red!70,very thick,minimum width=7.25mm,minimum height=7.25mm] {};

                \node at (1.75,1.75) [rectangle,draw=red!70,very thick,minimum width=7.25mm,minimum height=7.25mm] {};
 
     \node (p1) at (1.25,.85) {};
     \node (p2) at (1.65,1.85) {}; 
     \path (p1) edge[->,very thick,bend left,color=black] (p2); 
     
     \node (p3) at (1.2,.85) {};
     \node (p4) at (1.85,1.925) {}; 
     \path (p3) edge[->,very thick,color=black] (p4); 
     
     \node (p4) at (1.15,.9) {};
     \node (p5) at (1.8,1.65) {}; 
     \path (p4) edge[->,very thick,bend right,color=black] (p5); 
      
% Axes:

  %   \node (a1) at (0,-.5) {};
%     \node (a2) at (2.75,-.5) {};
 %    \path (a1) edge[->,thick] (a2);

\end{tikzpicture} 
}

\caption{Measuring how an outcome could be different when moving from one category to another may depend on how precisely one picks exemplars of the two categories. Given a two-dimensional feature space (given by $x$ and $y$ coordinates), suppose settings of those features are clustered into categories, as boxes in Subfigure (a). Shade of teal denotes likelihood of an outcome (say, interview status) for a given category. Subfigure (b) reveals a finer-grained picture of the space of individuals, at which level outcome status is binary (yes or no). Taking an individual from one category to another can be done in any number of different ways, and these different possibilities may also differ with respect to the outcome. To the extent that the alignment is causally consistent, this type of ambiguity will be absent.  %Two-dimensional feature space (X-axis is one feature, while the Y-axis is a second feature). Individuals (left) are coarse-grained (right) as a function of these two features. Teal is a third property (typically, an effect of X and Y) of interest (e.g., gets a call-back), and shade of teal in the coarse-grained picture is proportional to probability (frequency) of the property. A coarse-grained counterfactual (depicted in violet) may be ambiguous, such that, when the abstraction is causally inconsistent, some instantiations in the fine-grained picture (the middle arrow on the left) produce non-representative instances of the category with respect to the outcome.
} \label{fig:ambiguity}
\end{figure}

\new{This ambiguity does not arise if the alignment $\tau_{\text{SC}}$ is perfectly causally consistent, for then the effects of race correspond perfectly with those of each of its constituents. However,} as suggested in Section~\ref{sec:introducing abstraction}, once we relax the stringent requirement of causal consistency, and allow that the effect of race on interview status is some complex function of the varying effects of many different lower-level changes, we face the difficult task of aggregating those low-level contrasts into a single, higher-level one. (In the case of audit studies, this involves defining the equation that underlies the arrow $\textsf{Race}\rightarrow \textsf{Interview}$ in Figure~\ref{figure:abstraction model}, and thus deciding to what extent differential treatment on the basis of race-coded names fully captures discrimination on the basis of race.) This function, like the social constructionist alignment $\tau_{\text{SC}}$, will need to be informed by our understanding of the normative significance of discrimination. Similarly, even once we define the high-level structural function, whether an alignment is causally consistent ``enough'' is partly a normative question. To obtain a theory of race that is explanatory across a wider variety of contexts, or one that employs fewer high-level categories, we might have to relax causal consistency further. \new{It is an open question whether a metaphysically plausible alignment between race and its constituents can deliver cross-contextual causal consistency, even approximately.}\footnote{See Tolbert~\citeyearpar[p. 1105]{tolbertcausal} and \cite{tolbert2024restricted} for  skepticism that the effects of race are homogenous enough to lend racial categories explanatory power; in the framework of abstraction, this could lead to skepticism that there exists a cross-contextual, causally consistent alignment between race and its constituents.} \new{We underscore that in the abstraction framework, this is largely an empirical question: whether race is tenable as a cause just depends on whether it proves to be an explanatory way of clustering lower-level attributes. Current work in social science employs the framework developed here to test whether this is the case over %tists are even using the framework developed here to test whether this is the case, with 
real-world datasets [citation redacted for anonymous review].}%\cite{huang2025causalfeaturelearningsocial}.

%Recent work in the social sciences applies our framework to 
   
Suppose, however, that the problem of ambiguity is addressed, and that we have defined the effects of race as some complex function of the effects of its lower-level constituents. Then to the extent that alignment is causally consistent, we could rest assured that audit studies revealed an effect of race, i.e. that we acquire information about (\ref{eq:intervention on race}) via (\ref{eq:intervention on resume}):
\begin{align}
    &\mathbb{P}[\textsf{Interview Callback}\,|\, \text{do}(\text{Race}_1)] - \mathbb{P}[\textsf{Interview Callback}\,|\, \text{do}(\text{Race}_2)] \label{eq:intervention on race}\\
    %\approx \quad
    &\mathbb{P}[\textsf{Interview Callback}\,|\, \text{do}(\text{Resume}_1)] - \mathbb{P}[\textsf{Interview Callback}\,|\, \text{do}(\text{Resume}_2)]\label{eq:intervention on resume}.
\end{align}

\noindent In other words, audit studies would then provide information about the result of intervening on race, by intervening on resumes.

Even when this assumption is granted, there remains a second question about why the interventional difference (\ref{eq:intervention on resume}) is normatively relevant. For instance, in the resume audit study, the populations corresponding to the resumes used in the study may be \textit{atypical}, insofar as there is no guarantee that when one changes ``Greg'' to ``Jamal'' on a resume, one moves from a representative population of white people to a representative population of Black people (Figure~\ref{figure:atypicality}). In fact, either of these populations could be completely empty.\footnote{Tolbert~\citeyearpar[pp. 1100-3]{tolbertcausal} argues that because race and socioeconomic status are highly correlated (and indeed race may be defined partly in terms of the social stratification it sustains), race will violate the so-called ``positivity'' requirement, potentially undermining its status as a cause. In the framework of abstraction and applied to the audit study, this suggests the related worry that because race is correlated with socioeconomic status, and socioeconomic status is highly correlated with the information appearing on a resume, it will be difficult to find pairs of resumes which differ only in their race-coded names and which correspond to sufficiently large and representative populations.} If the resumes created by audit studies are highly atypical, it is unclear how they could shed light on a normatively salient population; the interventions performed by audit studies would simply be ``strange'' \citep[p. 10]{hu2020sex}.
\begin{figure}[H]
    \centering

  \begin{tikzpicture}[->, >=stealth,shorten >=1pt,auto,node distance=2cm,thick,scale=.9]
    \draw[thick] (0,0) ellipse (4 and 2);
    
    \draw[dotted, thick] (-2,0) ellipse (1.5 and 0.75);
    \draw[thick,fill=red!7.5] (-2,.25) ellipse (1 and 0.5);
    
    \draw[thick, smooth,fill=blue!7.5, opacity=0.6] plot [smooth cycle] coordinates {(2, 0.25) (2.5, 0.5) (2.3, 0.75) (2.5, 1) (2, 0.8) (1.5, 0.75) (1.7, 0.5) (1.5, 0.3) (2, 0.25)};
    \fill (-2, 3) circle (2pt);
    \fill (2, 3) circle (2pt);
    \draw[dotted,thick] (-2, .75) -- (-2, 3) node[midway,right, yshift=8pt] {\small $\tau_{\text{audit}}$};
    \draw[dotted,thick] (2, .75) -- (2, 3) node[midway,right, yshift=8pt] {\small $\tau_{\text{audit}}$};
    \node[left, align=right] at (-4, 1) {Population of individuals};
    \node[left, align=right] at (-4, 3.25) {Possible resumes};
    \draw[thick,fill=red!7.5] (-1.75, 3) rectangle (-0.25, 5);
    \draw[thick,fill=blue!7.5, opacity=0.6] (2.25, 3) rectangle (3.75, 5);
    \node at (-1, 4.5) {\dots};
    \node[scale=0.75] at (-1, 4.1) {Greg};
    \node at (-1, 3.7) {\dots};
    \node at (-1, 3.3) {\dots};
    \node at (3, 4.5) {\dots};
    \node[scale=0.75] at (3, 4.1) {Jamal};
    \node at (3, 3.7) {\dots};
    \node at (3, 3.3) {\dots};
    \draw[dotted, thick] (2, 0) ellipse (1.5 and 0.75);
    \draw[->, thick, bend right=20, line width=.5mm] (-2, 3) to (2, 3);
    \draw[->, thick, bend right=20, line width=.5mm] (-2, -.75) to (2, -.75) node[midway, below=12pt] {?};
\end{tikzpicture}
    
    \caption{Atypicality in the resume audit study. The resumes differ only in their names, but the resume for Greg corresponds to a large population of individuals (the red oval), who are representative of the larger group of people named Greg (the leftmost dotted oval), whereas the resume for Jamal corresponds to a small, atypical population of individuals (the blue splotch), rather than the larger group of people named Jamal (the rightmost dotted oval).}
    \label{figure:atypicality}
\end{figure}

Short of a requirement that we do not compare completely empty populations, normative relevance does not always require typicality. For instance, in the resume audit study, an auditor who believes Black students who attended majority-white schools should receive similar job opportunities as compared to white students who attended those schools might decide that the causal contrast between these populations is relevant, even if the former population is ``atypical.'' However, this choice of a causal contrast is based on a normative assumption that these populations deserve similar treatment, and the ability of an audit study to demonstrate discrimination depends on the strength of the argument for that assumption. Likewise, concluding from the \textit{absence} of such an interventional difference that there is not discrimination requires a normative assumption that there are no other contrasts that are normatively relevant---for instance, contrasts corresponding to populations reflective of differences in the schools typically attended by Black and white students.

There are thus at least two basic assumptions needed for audit studies to reveal whether discrimination has occurred:

% (KS) Suggested some for a second revision

\begin{enumerate}[label=(\roman*)]
    \item 
    There must exist an alignment between race and lower-level constituents, which is approximately causally consistent, such that testing effects of race-coded names provides information about the effects of race.
    \item 
    The interventional quantities revealed by audit studies must correspond to race subpopulations which are normatively relevant. For instance, a racial subpopulation  might be relevant because they are in some sense ``typical'' for that race; or they may be ``atypical'' but have features of normative relevance to expectations of similar treatment.  (To reveal the \textit{absence} of a causal effect of race, the interventional quantities revealed by audit studies would have to correspond to \textit{all} the racial subpopulations that are normatively relevant.\footnote{\new{We emphasize that many additional assumptions, often unwarranted, but unrelated to intervention identification, are needed to move from an audit's failure to confirm a causal effect of race or gender to a conclusion that there is not discrimination, because such a finding does not mean race or gender are not a cause of outcomes in some salient way. \cite{narayanan2022limits} outlines many obstacles to detecting these causal effects quantitatively, which can be illustrated by the resume audit study. For instance, a resume audit might be underpowered to detect very small differences in interview callback outcomes; yet these undetectable small differences can accumulate to large group disadvantages in the long run, leading to discrimination invisible in the ``snapshot'' dataset of an audit \cite[pp. 9-10]{narayanan2022limits}. Moreover, if most job discrimination occurs at the interview stage, then such discrimination will be invisible to the resume audits, which only looks at effects of race on callback rates \cite[pp. 13-15]{narayanan2022limits}.
While posing fundamental challenges to the use of audits to substantiate claims of the absence of discrimination, these challenges do not originate from puzzles about taking race or gender as a cause in principle, but rather from data availability and hypothesis formation about the points in a decision making process at which race or gender can act as a cause.  
}})
   
\end{enumerate}

Stated explicitly and precisely, these assumptions can be subjected to further normative and empirical challenges, and we make no claim to have defended them. These further debates target particular modeling assumptions, which can be stated within the framework of causal abstraction; they are not objections to causal models of discrimination as such.

\subsection{Intervention identification and experimental design}\label{subsec:experimental design}

We now illustrate how different responses to the above %intervention identification
challenges lead to differences in experimental design. Recall the in-person audit study, in which trained actors of different genders attend a job interview, presenting identical resumes and answering interview questions identically. An experimenter must determine whether to match or experimentally vary the appearances and mannerisms of actors of different genders---for instance, whether they are wearing a dress or a suit, whether or not they wear makeup, or how assertive they are.

When audit studies are modeled using the Single-Level Assumption, experimental designs must always be licensed by bespoke modularity assumptions. Any features of individuals that experimenters vary are treated as ``part of'' gender, while any feature they keep constant is treated as ``separate from'' gender, and placed in the causal diagram subject to a modularity assumption. When a variable is modular with respect to gender, it cannot be partially constitutive of it. Thus, different choices of intervention correspond to different assumptions about how gender is socially constructed.

This tight connection between assumptions about the social construction of gender and an audit study's choice of an intervention can lead to implausible and normatively undesirable conclusions about the constitution of gender, even by the auditor's own lights. Suppose an auditor designs a causal test of gender discrimination looking at differences in treatment between gender-nonconforming men and gender-conforming women. To vary gender-conformity across applicants, both applicants wear skirts. This implies, on the Single-level Assumption, that dress is modular with respect to gender. But this modularity assumption is strange in light of an underlying motivation of the experiment, namely that there are highly gendered standards about dress (which would suggest that dress partially constitutes gender). 

On the abstraction picture, by contrast, we can interpret the auditor's choice of intervention as committing to assumptions not just about (i) the  constitution of gender, but also about (ii) which interventions select normatively relevant populations for similar treatment. The auditor running a study about gender-nonconforming job applicants can include dress in their abstraction, corresponding to their understanding that standards of dress are gendered. They then select an intervention that compares women wearing skirts and men wearing skirts, corresponding to the normatively relevant comparison of treatment for gender-conforming women and gender-nonconforming men. Their choice of intervention is made \textit{in light} of how dress constitutes gender, rather than by excluding dress from the constitution of gender.

Of course, in the abstraction framework, it is still possible for the crux of an intervention identification to rest on what attributes constitute gender. \textit{Bostock vs. Clayton} provides one example. Should we vary the individual's sexual orientation to test for gender discrimination, or just vary biological sex and hold orientation fixed?\footnote{We put aside the question of whether discrimination on the basis of sexual orientation can be understood as a category additionally meriting protection from discrimination apart from sex. Our more limited aim is to articulate what distinguishes two proposals for counterfactual tests of gender discrimination in the abstraction framework, one similar to that of \cite{dembroff2020taylor}, and another one given in the dissenting opinion of \cite{bostock_clayton_county}. We do not analyze the majority decision of Bostock precisely on its own terms -- the opinion takes sex as biological, and consequentially, as \cite{dembroff2020taylor}, pp. 7-8, and \cite{kohler-hausmannSupremeConfusionCausality2022}, pp.84-86, point out, commits a modularity violation just like the dissent.}

The former approach, which varies the individual's sex but not that of their partner, does not require a distinction between gender conformity and non-conformity in the model of gender, and indeed the dissents in \cite{bostock_clayton_county} are motivated by the view that effects of sexual orientation do not constitute effects of gender.\footnote{For instance: ``The Court tries to convince readers that it is merely enforcing the terms of the statute [prohibiting sex discrimination], but that is preposterous.
Even as understood today, the concept of discrimination because of `sex' is different from discrimination because of
`sexual orientation.''' (\cite{bostock_clayton_county}, dissent of Alito p.3).} By contrast, on the latter approach, the high-level model in the abstraction might possess a single variable with four possible values: gender-conforming man, gender-conforming woman, gender-nonconforming man, and gender-nonconforming woman. The lower-level attribute of possessing a same-sex partner would then be aligned with gender-nonconformity, so varying sexual orientation is a way of intervening on gender (from gender-conforming man to gender-nonconforming man). Many social constructionist theories of gender indeed intend to highlight norms punishing non-conformity (cf. \citealt{kohler-hausmannSupremeConfusionCausality2022}, pp.88-89). Such modeling choices might be further justified, in part, by considerations of causal consistency; if gender-conforming and gender non-conforming men are in fact treated very differently, then without such a high-level distinction, the abstraction will not be very causally consistent, combining groups that are treated very differently together.%\footnote{Indeed, an analysis comparing the case of Bostock to a hypothetical one concerning gender conformity comes up in the majority opinion of \cite{bostock_clayton_county}, pp.22-23.}

In summary, the causal abstraction framework distinguishes a number of assumptions pivotal to the identification of an intervention. While the choice \textit{to} vary a certain attribute commits us to including it in the constitutive basis of a social kind (as with sexual orientation in \textit{Bostock vs. Clayton}), the choice \textit{not to} experimentally vary an attribute need not be motivated by its exclusion from the constitutive basis of the social kind (as with attire in the in-person audit study). In other words, being included in the constitutive basis for the relevant social kind is a necessary condition for being varied in an experimental design, but not a sufficient one; a feature may be held fixed, in order to ensure that the test targets normatively relevant subpopulations.

\section{Two notions of causal discrimination}\label{sec:two notions}

In the preceding sections, we have argued that the abstraction framework for modeling discrimination addresses the modularity problem, and that it provides the precision needed to distinguish several challenges for intervention identification. Some critics of causal approaches to discrimination have used both the modularity problem and challenges of intervention identification to motivate the suggestion that because race is socially constructed, discrimination claims are backed by constitutive explanation, rather than causal explanation (\citealt{hu2020sex}, p.10; \citealt{dembroff2020taylor}, p.9). In this section, we show that this suggestion can be made precise using causal abstraction as well; the appeal to constitutive explanation provides another interpretation of the causal notion of discrimination, not a non-causal alternative to it. We conclude that none of the worries we discuss are obstacles to causal models of discrimination as such.

To begin, suppose someone explains why a child, Oona, is not allowed to watch Game of Thrones by appealing to her membership in that social category---``because she is a child'' (\citealt{dembroff2020taylor}, p.9). \cite{dembroff2020taylor} point out that this sort of explanation operates ``at the level of social meanings and norms attached to the category \textit{child}, rather than at the individual level of Oona's age.'' There is an ambiguity in this observation, because there are two different causal contrasts by which social norms can perform explanatory work.

One kind of causal contrast is between two states of a socially embedded category. For instance, we might ask: Would Oona have been allowed to watch Game of Thrones were she an adult, rather than a child? To pose this counterfactual, we cannot just vary Oona's age---leaving all else constant---we must also vary features about Oona's social relations that would tend to change with her age, like the way Oona's parents interact with her. In the abstraction framework, this corresponds to selecting a low-level intervention on Oona's features, including her natural features and features about her social relations, that corresponds to a high-level intervention bringing her from ``child,'' to ``adult,'' and then measuring the causal effect of that intervention. Call this \textit{explanation by a socially constructed attribute.}

However, a second, relevant causal contrast concerns what the effect on the outcome would have been if the socially constructed category \textit{itself} had been constructed differently. For instance, we might ask: Would Oona have been allowed to watch Game of Thrones, if the social norms attending to childhood were different? On the abstraction framework, this causal contrast can be represented by actually changing the abstraction of the social category itself, so that, for instance, parental supervision of media consumption is \textit{not} partially constitutive of childhood; then, we measure the causal effect of this \textit{constitutive counterfactual} on the outcome of interest, in this case Oona's ability to watch Game of Thrones. Call this \textit{explanation by a social kind's constitutive norms.}

These two contrasts answer to different types of causal questions. The first is an assessment of the effect of a social kind \textit{as it is currently defined in our social structure}, while the second is about what sort of effect of a social kind would be possible \textit{if we reformed the social norms defining that kind}. These causal questions correspond to different ways of spelling out the notion of causal discrimination:

\begin{quote}
    \textit{Attribute Discrimination:} A person is directly racially discriminated against only if she is treated worse than others, and this is caused by her race, in the sense that were her race, and only her race, different, she would not have been treated worse than others.
\end{quote}

\begin{quote}
    \textit{Norm Discrimination:} A person is directly racially discriminated against only if she is treated worse than others, and this is caused by her race, in the sense that were race itself differently constituted, she would not have been treated worse than others.
\end{quote}

These should both be seen as live options. On the one hand, attribute discrimination is not a purely academic invention, gerrymandered to satisfy the demands of the causal inference framework or fit the constraints of modern antidiscrimination caselaw; audit studies originated in activist-led antidiscrimination efforts spanning back to the 1950s \citep{cherryMakingItCount2018}. On the other hand, the notion of norm discrimination is not unprecedented. For example, \cite{barocas-hardt-narayanan} propose that an individual is discriminated against if they would have had different opportunities absent structural inequality (cf. \citealt{liuReachFairness2024}). And as \cite{kohlerhausmann2018eddie} notes (p.1226), multiple legal theorists have proposed interpretations of antidiscrimination law as regulating against the pernicious impact of the social norms that constitute protected categories, suggesting the relevance of causal-constitutive counterfactuals to discrimination law.

It lies beyond the scope of this paper to adjudicate between these two notions of discrimination. However, we underscore that both are \textit{causal} notions of discrimination, both are amenable to social construction, and both can be formalized using causal abstraction. Indeed, in the abstraction framework, attribute discrimination measures an effect of a shift across the boundaries of the partition of lower-level attributes defined by the alignment. Meanwhile, norm discrimination requires that we first change the constitutive basis itself---the alignment of the high-level social kind to low-level attribute space---and then determine the causal profiles of the variables in the model, were that counterfactual alignment real (Figure~\ref{fig:abstraction-norms}).

\begin{figure}[H]
    \centering
    \begin{tikzpicture}[->, >=stealth,shorten >=1pt,auto,node distance=2cm,thick]
    
        % Top nodes
    \node[draw, circle] (Race) at (-4, 2) {\textsf{Race}};
    \node[draw, circle] (EffectTop) at (4, 2) {\textsf{Effect}};
    
    % Bottom nodes
    \node[draw, circle] (V1) at (-7, -2) {$V_1$};
    \node[draw, circle] (V2) at (-4, -2) {$V_2$};
    \node[draw, circle] (V3) at (-1, -2) {$V_3$};
    \node[draw, circle] (EffectBottom) at (4, -2) {\textsf{Effect}};

        % Causal relations
        \draw[->, thick] (V3) to (EffectBottom);
        \draw[->, thick, bend left=30] (V2) to (EffectBottom);
        \draw[->, thick] (Race) -- node[midway,above] {\textbf{?}}(EffectTop);

        % Abstraction relations
        \draw[dotted, thick] (V1) -- node[midway,left] {$\tau_{\text{ideal}}$?}  (Race);
        \draw[dotted, thick] (V2) --  (Race);
        \draw[dotted, thick] (V3) -- node[midway,right] {$\tau_{\text{actual}}$} (Race);
        \draw[dotted, thick] (EffectBottom) --  (EffectTop);
    \end{tikzpicture}
    \caption{Norm explanation in the abstraction model. While race is in fact constituted by $V_2$ and $V_3$ via $\tau_{\text{actual}}$, and thus has a certain effect, we can ask whether race would have the same effect, were it instead constituted by $V_1$ and $V_2$ via $\tau_{\text{ideal}}$. Above, a dotted arrow without a label specifies an abstraction relation common to both $\tau_{\text{actual}}$ and $\tau_{\text{ideal}}$. In general, $\tau_{\text{ideal}}$ and $\tau_{\text{actual}}$ can agree or disagree on the variables that belong to the constitutive basis of race, so long as they differentially align race with the values of these low-level variables.}
    \label{fig:abstraction-norms}
\end{figure}

\section{Conclusion}

    While pervasive, the notion of causal discrimination is difficult to spell out in a precise and plausible way. For the notion to make sense, it seems to require that race is an attribute of an individual sufficiently separable from their other attributes so that its causal role can be isolated. But if race is socially constructed, in what sense could it be separable in this way? In response, many have proposed that we give up on modeling race as causing worse treatment, either arguing that a distinct attribute (e.g. perception of race) is what causes worse treatment, or that (because causal models require modularity) attempts to causally model racial discrimination are fundamentally misguided.

    This paper has addressed the problem differently. We introduced a framework for reasoning about discrimination, in which race is a high-level abstraction of lower-level, in-principle manipulable features. In this framework, race can be modeled as itself causing worse treatment. The essential condition of modularity is ensured by allowing assumptions about social construction to be precisely and explicitly stated, via an alignment between race and its lower-level constituents. Such assumptions can then be subjected to further challenges. How do we define social kinds and their effects? Which population contrasts are normatively relevant to discrimination? More fundamentally: Are discrimination claims backed by causal explanations that contrast socially constructed \textit{attributes}, or by causal explanations that contrast \textit{the norms} by which these attributes are constructed? These questions have an empirical aspect and a normative one: they recommend different experimental designs, and their answers will depend in part on what makes discrimination wrongful. While answers to these questions lie beyond the scope of this paper, we have used the abstraction framework for modeling  discrimination to state them precisely, and to explore some of their potential implications. We conclude that these questions point to important avenues for further work at the intersection between ethics and philosophy of science, but not to any in-principle objections to causal models of discrimination as such.

\newpage

\subsection*{Acknowledgements} Many thanks to Hannah DeBrine, Myra Deng, Johann Frick, Elek Lane, Russell McIntosh, Christian Nakazawa, Naftali Weinberger, and Eliza Wells for comments and discussion, and to Caltech's Linde Center for Science, Society, and Policy (LCSSP) for supporting a research incubator to develop this paper. Special thanks to two anonymous reviewers from Noûs for extremely detailed and helpful feedback. Kara Schechtman was supported by a graduate fellowship award from Knight-Hennessy Scholars at Stanford University during initial work on this paper.

\bibliographystyle{apalike}
\bibliography{refs}{}

\end{document}